\UseRawInputEncoding

\documentclass[lettersize,journal]{IEEEtran}
\usepackage{amsmath,amsfonts}
\usepackage{algorithmic}
\usepackage{algorithm}
\usepackage{array}
\usepackage[caption=false,font=normalsize,labelfont=sf,textfont=sf]{subfig}
\usepackage[font=normalsize,labelfont=sf,textfont=sf]{caption}
\usepackage{textcomp}
\usepackage{stfloats}
\usepackage{url}
\usepackage{verbatim}
\usepackage{graphicx}
\usepackage{cite}
\usepackage{tabularx}  % 引入tabularx包
\usepackage{lipsum}    % 引入lipsum包生成示例文本
\usepackage{array}     % 自定义列类型
\usepackage{booktabs}  % 专业表格线
\usepackage{ragged2e}  % 对齐控制
\usepackage{colortbl}
\usepackage{color}
\usepackage{xcolor}    % 定义新的列类型 Y，支持按权重分配宽度
\usepackage{hyperref}

\newcolumntype{Y}{>{\raggedright\arraybackslash}X}

\begin{document}

\title{A Comprehensive Survey of Electrical Stimulation Haptic Feedback in Human-Computer Interaction}

\author{
Simin Yang \thanks{\textbf{Simin Yang} is with the Department of the Division of Emerging Interdisciplinary Areas, Hong Kong University of Science and Technology, Hong Kong, China (e-mail:syangcj@connect.ust.hk).}, 
Xian Wang \thanks{\textbf{Xian Wang} is with the Department of Industrial and Systems Engineering, Hong Kong Polytechnic University, Hong Kong, China (xiann.wang@connect.polyu.hk).}, 
Yang Li \thanks{\textbf{Yang Li} is with the Department of Mechanical and Aerospace Engineering, Hong Kong University of Science and Technology, Hong Kong, China,(e-mail: ylink@connect.ust.hk).}, 
Lik-Hang Lee \thanks{\textbf{Lik-Hang Lee} is with the Department of Industrial and Systems Engineering, Hong Kong Polytechnic University, Hong Kong, China (lik-hang.lee@polyu.edu.hk).}, 
Tristan Camille BRAUD \thanks{\textbf{Tristan Camille BRAUD} is with the Department of Computer Science and Engineering, Hong Kong University of Science and Technology, Hong Kong, China (braudt@ust.hk).}, 
Pan Hui \thanks{\textbf{Pan Hui} is with the Department of the Division of Emerging Interdisciplinary Areas, Hong Kong University of Science and Technology, Hong Kong, China, and also with Department of Computer Science, University of Helsinki, 00014 Helsinki, Finland (e-mail: panhui@cse.ust.hk).}
\thanks{This research was supported in part by a grant from the Guangzhou Municipal Nansha District Science and Technology Bureau under Contract No.2022ZD01 and the MetaHKUST project from the Hong Kong University of Science and Technology (Guangzhou).}
}

\maketitle
\begin{abstract}
Haptic perception and feedback play a pivotal role in interactive experiences, forming an essential component of human-computer interaction (HCI). In recent years, the field of haptic interaction has witnessed significant advancements, particularly in the area of electrical haptic feedback, driving innovation across various domains. To gain a comprehensive understanding of the current state of research and the latest developments in electrical haptic interaction, this study systematically reviews the literature in this area. Our investigation covers key aspects including haptic devices, haptic perception mechanisms, the comparison and integration of electrical haptic feedback with other feedback modalities, and their diverse applications. Specifically, we conduct a systematic analysis of 110 research papers to explore the forefront of electrical haptic feedback, providing insights into its latest trends, challenges, and future directions.
\end{abstract}


\section{Introduction}

Touch is one of the most fundamental human senses, responsible for conveying critical information about the shape, texture, hardness, and temperature of objects. In the field of human-computer interaction (HCI), touch works synergistically with other senses, such as vision and hearing, to offer users a more natural and immersive experience. In recent years, the rapid advancements in virtual reality (VR), augmented reality (AR), and the metaverse have further underscored the essential role of touch in these immersive environments. Particularly in the context of the metaverse, touch has transitioned from being a speculative concept to a tangible requirement for enabling users to interact with complex and immersive virtual environments \cite{dionisio20133d, weinberger2022metaverse, ritterbusch2023defining, lee2024all}. However, compared to the technological maturity of vision and auditory feedback systems, the research and practical applications of haptic feedback, especially tactile feedback, remain in a relatively nascent stage.

% discuss the haptic sensation is important
Although vision and audition can already allow people to process spatial information, touch, based on contact interaction, enables people to obtain the surface properties of objects and has good spatio-temporal discrimination \cite{zhou2022electrotactile, lederman2009haptic}. 
Integrating 3D spatial perception and haptic feedback into interfaces can help users achieve a sense of immersion in virtual environments, thereby enhancing interactivity \cite{magnenat2006haptics}. Tactile perception plays an indispensable role in how humans interact with their environment, offering not just the ability to touch, but to feel and manipulate objects, which is crucial for realistic and engaging virtual interactions. Current implementations of tactile feedback in the Metaverse are largely simplistic, often limited to vibrations or force feedback \cite{kurzweg2024survey}, which do not replicate the rich, nuanced sensations experienced in the real world. 

There are various technology solutions for haptic feedback, including vibration devices, skin deformation, and mid-air devices \cite{bermejo2021survey}. Electrotactile stimulation is one of the haptic feedback technology methods, which was proposed by Saunders \cite{zhou2022electrotactile, saunders1983information}. The principle involves applying currents or pulses on the skin surface. Once the skin's nerves receive this electrical input, they relay signals that the brain interprets as tactile sensations. By adjusting the parameters of the electrical current, it is possible to simulate various types of touch. Electrical tactile feedback offers distinct advantages over the more established vibration feedback and pneumatic solutions: (1) The electrodes used in electrical tactile feedback can be made much smaller, significantly reducing the size of wearable devices \cite{bermejo2021survey}; (2) Electrical tactile feedback does not require direct contact since the electrical signals can propagate through the body, allowing tactile sensations to be experienced even without wearing the device directly on the targeted area.

Extensive research in human-computer interaction (HCI) has discussed the enhancement of tactile experiences through electrical feedback. Yet, there is still a lack of comprehensive studies addressing the full-body application of this technology. This study aims to thoroughly examine: (1) the current technological solutions for electro-haptic feedback; (2) the advantages and limitations of existing electrical feedback approaches; (3) the industries where electrical feedback is applied; and (4) the potential future developments of electrical feedback technology.

\subsection{Human Haptic Perception}
Haptic perception occurs within the human skin. Human skin is commonly categorized into three types: hairy skin, hairless skin (glabrous, such as palms and soles), and mucous skin (found within the mouth and other bodily openings) \cite{goodwin2008physiological, visell2009tactile}. Hairy skin is the most prevalent type, accounting for 95\% of the skin on the human body, and it is characterized by the presence of hair follicles \cite{magnenat2006haptics}. In hairy skin, the epidermis is less than 0.1 millimeters thick, and the dermis is 1–2 millimeters thick. Hairless skin, primarily found on the palms and soles, is thicker than hairy skin; its epidermis is approximately 1.5 millimeters thick, and the dermis extends to a depth of about 3 millimeters \cite{gould2018superpowered}. 

Haptic perception often involves the integration of two subsystems: the cutaneous (or tactile) and kinesthetic systems  \cite{lederman2009haptic, grunwald2008human}. Cutaneous receptors located beneath the skin, namely mechanoreceptors, thermoreceptors, and nociceptors \cite{marzvanyan2019physiology}. Mechanoreceptors are responsible for sensing physical changes, including touch, pressure, vibration, and stretch \cite{marzvanyan2019physiology}. 
Thermoreceptors are specialized for the detection of temperature variations, while nociceptors are dedicated to the perception of pain. These receptors are located in various types of skin, including glabrous skin, hairy skin, and mucous skin \cite{goodwin2008physiological, visell2009tactile}. Kinesthesia, also known as proprioception, is induced by mechanoreceptors in muscles, skin, and joints, as well as from central signals associated with movement \cite{taylor2022kinesthetic}. 
It is utilized not only to assess the actions of one's own body but also to determine the properties of objects being interacted with, such as weight and hardness \cite{taylor2022kinesthetic}. 

In mammals, there are six types of innocuous-touch receptors: Merkel cell-neurite complexes, Meissner corpuscles, Pacinian corpuscles, Ruffini receptors, hair follicles, and free nerve endings \cite{el2011haptics}. Hair follicles can detect light touch. Free nerve endings are the simplest type of sensory receptor within the skin. Despite their small size, they possess distinct characteristics. The skin is organized into a mosaic of territories, each consisting of free nerve endings that exhibit heightened sensitivity to distinct stimuli like heat, cold, touch, or pain \cite{carlson2018human}. Each territory responds exclusively to one type of stimulus. For example, a territory dominated by ``pain'' free nerve endings may also react to cold; however, it primarily transmits pain signals to the brain \cite{carlson2018human}.

The mechanoreceptors located in glabrous skin include Pacinian corpuscles, Ruffini endings, Meissner corpuscles, and Merkel’s discs (detailed characteristics available in \autoref{tab:functions_Of_mechanoreceptors}). Additionally, Merkel cells are also found in hairy skin \cite{Abraira2013}. Meissner's corpuscles and Merkel discs are characterized by small and localized receptive fields, with mean receptive areas of approximately 13 mm² and 11 mm², respectively. This configuration enables them to precisely detect the location of touch and subtle textural changes, making them ideally suited for processing high-resolution tactile information. Conversely, Pacinian corpuscles and Ruffini endings possess large and less localized receptive fields, with mean receptive areas of 101 mm² and 59 mm², respectively. This allows them to perceive stimuli over broader areas, albeit with lower resolution for precise location and detailed discrimination \cite{el2011human}. In addition to their receptive fields, the four classes of mechanoreceptors can be divided into slow-adapting and fast-adapting types based on their rate of adaptation \cite{lederman2009haptic}.

\begin{table*}[ht]
    \centering
    % \small
    \begin{tabularx}{\textwidth}{ 
        >{\hsize=0.10\textwidth\raggedright\arraybackslash}X |  % receptor
        >{\hsize=0.28\textwidth\raggedright\arraybackslash}X |  % characteristics
        >{\hsize=0.13\textwidth\raggedright\arraybackslash}X |  % perception
        >{\hsize=0.24\textwidth\raggedright\arraybackslash}X |  % location
        >{\RaggedRight}X                                        % frequency
    }
    \hline
    % \rowcolor[rgb]{0.851,0.851,0.851}
    \textbf{Receptor} & 
    \textbf{Response Characteristics} &
    \textbf{Perception} &
    \textbf{Location} &
    \textbf{Frequency} \\ 
    \hline
    Merkel   & Slow-adapting type I (SA I)   & Indentation   & Basal Layer of Epidermis & 0.4-100 Hz \\
    Meissner & Fast-adapting type I (FA I)   & Movement      & Dermal Papillae          & 10-200 Hz  \\
    Pacinian & Fast-adapting type II (FA II) & Vibration     & Deep Dermis              & 40-800 Hz  \\
    Ruffini  & Slow-adapting type II (SA II) & Stretch       & Dermis                   & 7 Hz       \\
    \hline
    \end{tabularx}
    \caption{Sensory innocuous-touch receptors in mammals.}
    \label{tab:functions_Of_mechanoreceptors}
\end{table*}

Proprioceptive (kinesthetic) receptors are classified into four principal types: Golgi-type endings located within joint ligaments, Ruffini endings situated in joint capsules, Golgi tendon organs embedded in muscles, and Muscle Spindles also found within muscular tissue \cite{hale2004deriving}. These receptors collaboratively function to sense the position and movement of the limbs, thereby providing critical feedback for the coordination of motor activities.

The term electrotactile is currently the most commonly used term to describe electrical haptic feedback. However, based on literature related to human perception, we note that haptics encompasses both tactile and kinesthetic sensations. In our study, we observed that certain electrical stimuli can induce muscle contractions, which objectively result in kinesthetic sensations or pseudo-kinesthetic effects. Despite this, previous studies have not rigorously defined these terms. In this paper, we specifically discuss studies that involve electrical stimulation causing kinesthetic or pseudo-kinesthetic sensations, providing a more structured perspective on this aspect of electrotactile feedback.

\subsection{Electrical Stimulation Model}
Studies on electrical stimulation were already underway by the 1970s \cite{triggs1974some}. The principal mechanism of electrotactile feedback involves simulating tactile sensations by controlling the intensity, frequency, and waveform of electric currents. Electrodes serve as actuators, transmitting mild electrical currents that directly stimulate the sensory nerves in the skin. This stimulation mimics the neural activity induced by physical contact, thereby eliciting the illusion of tactile sensations in the brain. Compared with the traditional vibration feedback actuators, electrotactile feedback actuators often possess advantages such as being lightweight, easily deformable, low in weight, energy-efficient, and mechanically robust \cite{kajimoto2016electro}.

\begin{figure}[h]
    \centering
    \includegraphics[width=\linewidth]{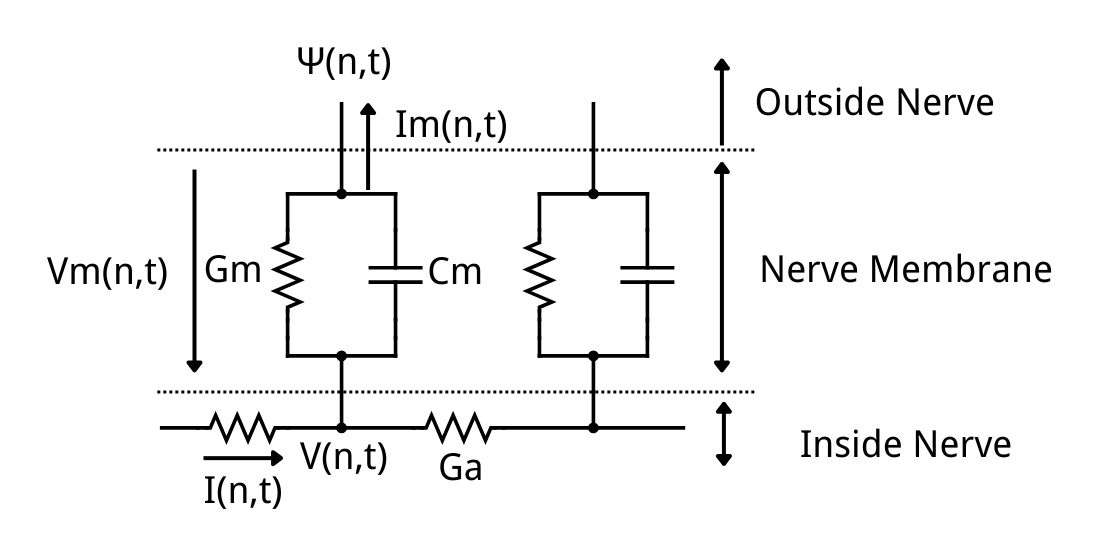}
    \caption{Electrical representation of electrical stimulation. Note that the current direction flows from the interior of the nerve to the exterior. This occurs because the depolarization process inside the nerve results in the accumulation of a large number of positive charges, leading to a current that moves from the inner region of the nerve to its outer region.}
    \label{fig:ElectircModel-1}
\end{figure}

In Kajimoto's study \cite{kajimoto2016electro}, Rattay's bioelectric model \cite{rattay1993modeling} was utilized to explain the principles of electrotactile feedback, and a model of the activation function was proposed:

\begin{equation}
    -G_a\frac{d^2}{dx^2}V_m + C_m\frac{d}{dt}V_m + C_mV_m = G_a\frac{d^2}{dx^2}\Psi
    \label{eq: activation function}
\end{equation}

Here, $G_a$ represents the internal conductance, $V_m$ denotes the membrane voltage difference caused by the electrical stimulation, $C_m$ refers to the curved membrane capacitance, $G_m$ indicates the conductance of the nerve membrane, and $\Psi$ represents the external membrane potential (See at \autoref{fig:ElectircModel-1}). The \autoref{eq: activation function} represents the activation function, where the term on the right side denotes the intensity of electrical stimulation, and the term on the left side represents the voltage differential generated as a result of the electrical stimulation. The transmembrane potential is formed by the current generated from one or multiple electrodes. Based on this activation function, several fundamental design principles can be identified: (1) If the nerve is parallel to the skin surface, only cathodic currents can induce nerve depolarization. Conversely, nerve perpendicular to the skin surface requires anodic current stimulation. (2) Deeper nerves are more difficult to activate. (3) The activation functions of multiple electrodes may cancel each other out. As a result, a more considerable distance between electrodes leads to more substantial and deeper stimulation.

In addition to the spatial distribution, the temporal distribution must also be considered. A typical waveform is the first-order input of a rectangular waveform (when $t \geq 0$, $\psi(t) = -V$; when $t < 0 $, $\psi(t) = 0$). According to Kajimoto's study \cite{kajimoto2016electro}, the step response of the membrane voltage difference and the membrane voltage difference threshold are given by the following formula:

\begin{equation}
    V_{th} \leq V_m = \frac{G_a}{G_a + G_m}(1-exp(-\frac{G_a + G_m}{C_m}T))V
    \label{eq: step response}
\end{equation}

Therefore, when the nerve is just stimulated, the relationship between the pulse width $T$ and the pulse amplitude $V$ is as follows:

\begin{equation}
    V = \frac{1}{b} \frac{V_{th}}{1-exp(-aT)}
    \label{eq: voltage and threshold}
\end{equation}

Here, $a = (G_a + G_m)/C_m$ and $b = G_a/(G_a + G_m)$. The \autoref{eq: voltage and threshold} demonstrates that the threshold pulse amplitude is inversely proportional to the pulse width, and that higher current results in greater charging efficiency. The parameters $G_a$, $G_m$, and $C_m$ are intrinsic properties of the nerve. Specifically, $G_a$ increases as the cross-sectional area of the nerve becomes larger, while $G_m$ and $C_m$ increase with the perimeter of the nerve. Consequently, thicker nerves exhibit a faster rise in $V_m$ and have lower activation thresholds. However, this threshold advantage diminishes as the pulse width increases. Since pain-related nerves are generally thinner than mechanoreceptor-related nerves, narrower pulses are typically used for stimulation.

\subsection{Electrical Stimulation}
In electro-haptic feedback devices, electrical stimulation is delivered to the body through electrodes. When the skin's nerves receive the electrical current, the brain generates a haptic sensation. Several factors contribute to this haptic feedback, including the hardware circuitry related to the electrodes, the properties of the skin, and various electrical parameters such as waveform, amplitude, frequency, and pulse width. 

\subsubsection{Hardware and stimulation modes design}
The hardware for electrical stimulation fundamentally relies on components such as a power supply, microprocessor, digital-to-analog converter (DAC), switching circuits, and electrodes. The stimulation mode can be divided into three modes: current mode stimulation, voltage mode stimulation, and switched capacitor (charge-based) mode of stimulation \cite{banerjee2020energy}. Current-mode stimulation, also referred to as constant current stimulation (CCS), offers the advantage of enabling relatively straightforward control of the current. However, its primary drawback is the limited voltage compliance. Voltage-mode stimulation, on the other hand, applies a constant voltage across the tissue, with the delivered current being dependent on the impedance of the electrodes and the tissue. Due to the nonlinear and time-varying properties of both electrodes and tissue, the current delivered in voltage-mode stimulation is not constant. Switched Capacitor (Charge-Based) Mode of Stimulation introduces a set of capacitors and employs an orthogonal clock to enable the charging and discharging process, thus achieving circuit-switching functionality. Its primary advantage lies in addressing the issue of charge balancing.

\subsubsection{Electrical parameters}
Common electrical stimulation parameters include pulse width, pulse frequency, duty cycle, stimulation intensity, and waveform type, all of which play a crucial role in defining the characteristics and effects of electrical stimulation. Common electrical stimulation waveforms include square pulse, sawtooth (sawtooth wave and reverse sawtooth), and sine waves. Among these, the pulse waveform is the most commonly used.

Pulse width, also referred to as pulse duration, means the time duration of a signal pulse, typically measured in microseconds ($\mu\text{s}$). The pulse width can be categorized into three distinct ranges: short pulse width, defined as less than 200 $\mu\text{s}$; medium pulse width, ranging from 200 $\mu\text{s}$ to 700 $\mu\text{s}$; and long pulse width, characterized as greater than 700 $\mu\text{s}$.

Pulse Frequency ($f$) means the number of pulses delivered per second, measured in Hertz ($Hz$) or Pulses Per Second ($PPS$). It is inversely related to the signal period ($T$), with $T = 1/f$. The duty cycle ($D$) is defined as the proportion of time the stimulation is "on" relative to the total cycle time, expressed as a percentage ($\%$), given by $D = \text{Pulse Width} / \text{Period}$.

Stimulation intensity corresponds to the amplitude of the stimulation signal, which can be measured in either milliamperes ($mA$) for current or volts ($V$) for voltage. The waveform type is the shape of the stimulation signal, which can be divided into square wave, biphasic wave, monophasic wave, sine wave, triangular wave, and so on. 

\subsection{Related Survey}
A substantial body of survey research exists on haptics, addressing a wide variety of topics. For instance, Culbertson et al. \cite{culbertson2018haptics} provided a comprehensive examination of methods for simulating tactile perception. Xia et al. \cite{xia2018new} explored five types of haptic rendering techniques: those for rigid-rigid interactions, rigid-deformable interactions, rigid-fluid interactions, image- and video-based interactions, and texture-based haptic rendering. Hamza et al. \cite{hamza2019haptic} focused on interfaces for haptic feedback devices, particularly those utilizing force-feedback and vibrotactile hardware. Further research on haptic feedback devices has expanded to include wearable haptic devices \cite{adilkhanov2022haptic} and encounter-type haptic devices \cite{mercado2021haptics}.

In addition to these works, other surveys have concentrated on specific subfields of haptics. For example, hand-based haptics have been systematically reviewed in several surveys \cite{tong2023survey, caeiro2021systematic, wang2018toward, dipietro2008survey, pacchierotti2017wearable}. Non-contact haptic interactions have also been extensively studied \cite{rakkolainen2019survey, rakkolainen2020survey, arafsha2015contactless, tsalamlal2013non}. Surveys on the integration of haptics with networks have examined topics such as tactile internet and networked haptic systems \cite{emami2024survey, tychola2023tactile, promwongsa2020comprehensive, tan2020methodology, antonakoglou2018toward, huisman2017social}. Haptics and its relationship with emotions have been explored in works on affective and emotional haptic design \cite{vyas2023descriptive, maclean2022designing, eid2015affective}. Other notable areas of investigation include soft haptics \cite{zhu2022soft, yin2021wearable} and sensory illusions \cite{kurzweg2024survey, ujitoko2021survey}, among others.

The exploration of electrical haptic feedback remains relatively limited. Zhou et al. \cite{zhou2022electrotactile} investigated the perceptual characteristics of the skin, the evaluation methods of electrical haptic feedback, and discussed its applications in various scenarios. Similarly, Kourtesis et al. \cite{kourtesis2022electrotactile} focused on the application of electrical haptic feedback for the hand and arm. Electrical haptic feedback, with its inherent advantages such as ultra-thin and lightweight designs and validated haptic rendering capabilities, holds significant potential for delivering full-body tactile feedback. This makes it a highly promising technology in the field of human-computer interaction. To fully realize this potential, we aim to conduct a comprehensive study that explores the underlying technologies, investigates the perceptual mechanisms of electrical haptic feedback, and examines both current advancements and future trends to enhance user haptic experiences.

\section{Method}
The objective of this survey is to comprehensively review existing research in the field of Human-Computer Interaction (HCI) that focuses on the application of electrical haptic feedback in virtual environments. To address our research questions, we adhere to the PRISMA 2020 guidelines \cite{page2021prisma} for systematic reviews. The PRISMA flow diagram is presented in \autoref{fig:PRISMA}.

\subsection{Research Questions}
Our research question is: How can electrical haptic feedback be utilized to enhance user experience in the virtual world? To better address our research question, we have divided it into the following sub-questions:
\begin{enumerate}
    \item RQ1: What is the current state of electrical haptic feedback devices?
    \item RQ2: What types of haptic perceptions are induced by electrical stimulation?
    \item RQ3: How does electrical haptic perception enhance the user experience?
    \item RQ4: What are the advantages and limitations of electrical haptic feedback, and what are the prospects for its future development?
\end{enumerate}

\subsection{Search Strategy}

According to the h-index provided by Google Scholar Metrics\footnote{\url{https://scholar.google.de/citations?view_op=top_venues&hl=en&vq=eng_humancomputerinteraction}}(accessed on December 10, 2024), the top 20 publications in the field of Human-Computer Interaction (HCI) primarily accessible through ACM Digital Library \footnote{\url{https://dl.acm.org/}} and IEEE Xplore Digital Library \footnote{\url{https://ieeexplore.ieee.org/}}. Therefore, we selected these two databases as the primary sources for our data collection. In addition, we conducted a preliminary search on Google Scholar and identified several publications that are relevant to research on electrical haptic feedback. As a result, we included the following publications in our study: Nature Communications, Nature Electronics, npj Flexible Electronics, Nature Machine Intelligence, Science Advances, and Science Robotics.

For keyword selection, we explored terms such as “electrotactile,” “electrotactile,” “electro tactile,” “electrocutaneous,” “electric stimulation,” and “electric feedback.” Using these keywords, we searched the ACM Digital Library (2014–2024) with results set to appear “anywhere” in the article and filtered for research articles only. This yielded 144 results, which were manually screened by title and abstract to exclude irrelevant articles. If new relevant terms emerged during this process, they were added to the keyword set and used for additional searches. The finalized keyword set included:

\begin{center}
    \parbox{0.9\linewidth}{``electrotactile'' OR ``electro tactile'' OR ``electrocutaneous'' OR ``electro-cutaneous'' OR ``electro cutaneous'' OR ``electrotactile stimulation'' OR ``electrical tactile feedback'' OR ``electric haptic feedback'' OR ``electrical feedback'' OR ``electric stimulation'' OR ``electrical muscle stimulation'' OR ``electric tactile feedback''}
\end{center}

We then conducted searches across databases. Articles were included if any keyword appeared in the title, abstract, or keyword section. In the ACM Digital Library, we filtered for research articles, and in IEEE Xplore, we included only conference papers and journal articles. 

\subsection{Eligibility}
\label{section:eligibility} 
In this study, we focused exclusively on research articles. Our include criteria are as follows: 
\begin{enumerate}
    \item The study must focus on electrical haptic feedback technology, specifically targeting human users rather than machines.

    \item The applications of haptic feedback should be directly related to HCI, with the primary goal of enhancing haptic experiences for users. Applications in the medical domain are beyond the scope of this study.

    \item The study should explore how electrical haptic feedback enhances user experience. 

\end{enumerate}
Criterion (1) requires the articles to focus on electrotactile feedback technology. This includes studies on the application of electrotactile techniques, innovations in hardware circuits, or the design of electrotactile rendering methods. Exclusion criteria are articles that primarily discuss other haptic technologies, such as vibrotactile feedback, ultrasonic haptics, force feedback, or thermal feedback, without addressing electrotactile feedback.

Criterion (2) requires that the studies involve electrical haptic interactions designed to elicit tactile sensory experiences in humans. Exclusion criteria include articles that focus on specialized medical applications or tactile experiences not applied directly to the human body.

Criterion (3) requires that the articles must explore user experience, including user testing and feedback. Exclusion criteria are studies that do not address user experience, such as those focusing solely on technical implementations or device development, or those that lack experimental or user data and remain purely theoretical.

\subsection{Screening Process}

\begin{figure*}[htbp]
    \centering
    \includegraphics[width=.7\linewidth]{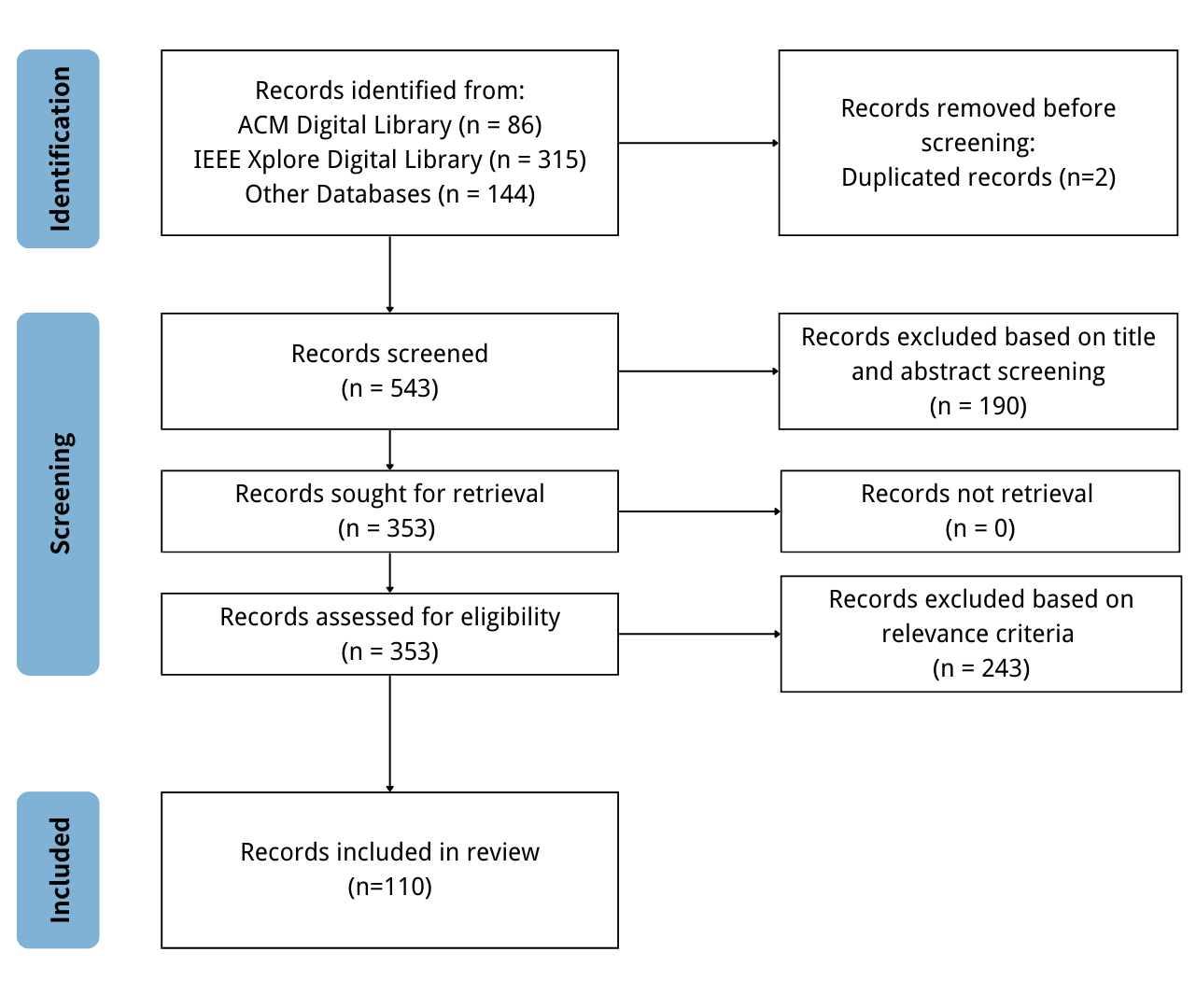}
    \caption{PRISMA flow diagram showing the stages of identification, screening, and inclusion of our systematic search.}
    \label{fig:PRISMA}
\end{figure*}

Prior to screening, two duplicate records from all search results were removed. Records were then screened in two phases. All screening records were independently evaluated by two researchers. Articles deemed unsuitable by both researchers were excluded, while those considered suitable proceeded to the next stage. Articles where consensus cannot be reached were retained for further evaluation in subsequent phases. First, the titles and abstracts of all remaining 543 records were reviewed against the eligibility criteria presented in \ref{section:eligibility} . A total of 190 records were excluded from this step. The remaining 353 records were then reviewed again in a second phase, this time using the full text. A further 243 records were excluded at this stage, leaving 110 records in our final corpus. The complete PRISMA flow can be seen in \autoref{fig:PRISMA}.

\subsection{Data Extraction}
To extract relevant information from the included articles, we developed a data extraction rubric that enables us to systematically capture the key parameters of electric haptic feedback. Initially, we randomly selected 10 articles to evaluate their content and to draft a preliminary extraction rubric. This rubric was then refined and validated using another set of 10 articles, resulting in the final version of the data extraction rubric in \autoref{tab:data extraction}. \textbf{DE1} and \textbf{DE2} pertain to the general description of the articles, where we extracted metadata such as authors, publication year, and journal information using DOI. \textbf{DE3} focuses on the research objective of the article, which, together with \textbf{DE4}, identifies the primary scope and direction of the study. \textbf{DE5} specifies the location of electrical stimulation. \textbf{DE6} captures the tactile perceptions elicited in users. \textbf{DE7} and \textbf{DE8} describe the types of electrical stimulation devices used. \textbf{DE9-DE12} document the parameters of electrical stimulation, including waveform, frequency, bandwidth, and intensity. \textbf{DE13} provides details about the participants involved in the experiments, such as sample size, age distribution, and gender ratio. \textbf{DE14} describes the experimental methods employed in the studies, such as perception-based experiments or task-oriented experiments. Finally, \textbf{DE15} summarizes the main conclusions of each article.

\begin{table*}[htbp]
\centering
\small
\caption{Data Extraction Rubric.}
\label{tab:data extraction}
\begin{tabularx}{\textwidth}{
    >{\RaggedRight}p{0.5cm}    % ID
    >{\RaggedRight}p{3.0cm}    % Data Extraction
    >{\RaggedRight}X         % Type
}
\toprule
\rowcolor[rgb]{0.851,0.851,0.851}
\textbf{ID} & 
\textbf{Data Extraction} & 
\textbf{Type} \\
\midrule
DE1  & DOI       & Open Text \\
DE2  & Title     & Open Text \\
DE3  & Objective & Open Text \\
DE4  & Research Classification & 
Hardware Research, Perception Research, Multisensory Research, Application Research \\
DE5  & Stimulation Location & 
Face, Forehead, Mouth, Tongue, Earlobe, Neck, Shoulder, Upper Arm, Elbow, Forearm, Wrist, Palm, Back of Hand, Finger, Fingertip, Thumb, Index Finger, Middle Finger, Thigh, Lower Leg, Ankle, Feet, Other \\
DE6  & Sensation & Tactile, Kinesthetic, Multisensory, Other \\
DE7  & Device Type & Single-Pair Electrode, Electrode Array \\
DE8  & Electrode Type & Metal Electrodes, Conductive Material Electrodes, Flexible Electrodes, Hard Electrodes, Unknown, Other \\
DE9  & Wave Type & Square Pulse, Sawtooth, Sine, Monophasic Pulse, Biphasic Pulse, Other \\ 
DE10 & Stimulation Frequency & Open Text \\
DE11 & Stimulation Duration  & Open Text \\
DE12 & Stimulation Amplitude & Open Text \\
DE13 & Participants & Open Text \\
DE14 & Experiment Methods & Open Text \\
DE15 & Key Findings & Open Text \\

\bottomrule
\end{tabularx}
\end{table*}

\section{Result}
\subsection{Overview of Included Publications}
\begin{figure*}[h]
    \centering
    \includegraphics[width=\linewidth]{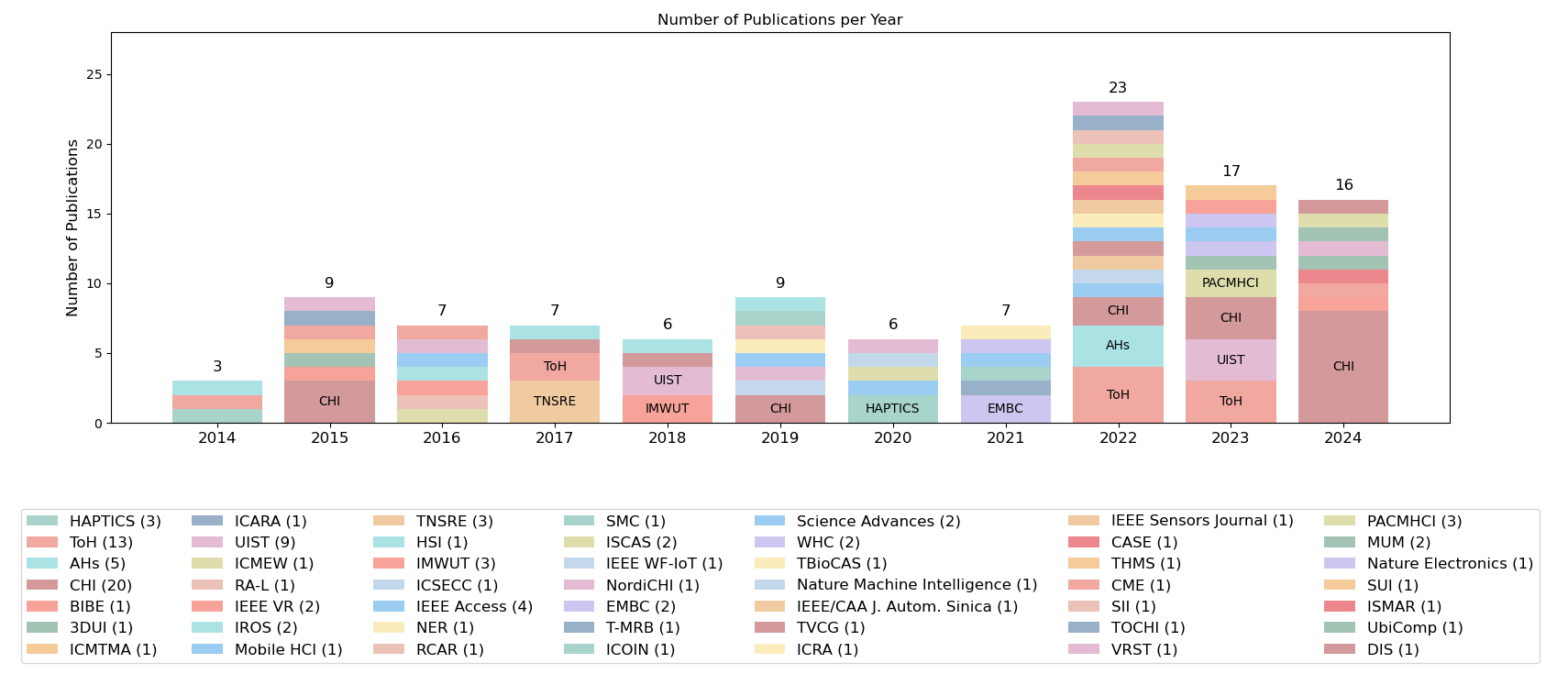}
    \caption{Publications Per Year (N=110). The chart marks the journals/conferences where the annual number of publications is greater than or equal to 2. The numbers in the caption represent the total number of publications for each journal/conference.}
    \label{fig:publications_peryear}
\end{figure*}

We meticulously extracted pertinent information from the 110 publications identified during our screening. The selected corpus covers publications from 2014 to 2024, with annual publication frequencies detailed in \autoref{fig:publications_peryear}. These works were predominantly disseminated through leading scientific publishers, including the American Association for the Advancement of Science (AAAS), Springer Science and Business Media LLC, the Institute of Electrical and Electronics Engineers (IEEE), and the Association for Computing Machinery (ACM). We found that the Conference on Human Factors in Computing Systems\footnote{\url{https://dl-acm-org.lib.ezproxy.hkust.edu.hk/conference/chi}} (CHI) remains the most prominent venue for research on haptic feedback (n=20), followed by the IEEE Transactions on Haptics\footnote{\url{https://ieeexplore.ieee.org/xpl/RecentIssue.jsp?punumber=4543165}} (ToH) (n=13). Additionally, the User Interface Software and Technology\footnote{\url{https://dl-acm-org.lib.ezproxy.hkust.edu.hk/conference/uist}} (UIST) conference, which focuses on user interface hardware and technologies, also includes discussions on user interfaces involving electrical stimulation feedback (n=9). 

\begin{table*}[htbp]
\centering
\caption{Distribution of the Electrical Haptic Publications. }
\label{tab:papers}
\begin{tabularx}{\textwidth}{
    >{\RaggedRight\arraybackslash}p{1.8cm}  % Location列：固定宽度+自动换行
    >{\RaggedRight\arraybackslash}p{3.3cm}  % Device
    >{\RaggedRight\arraybackslash}p{3.3cm}  % Principles and Perception
    >{\RaggedRight\arraybackslash}p{3.3cm}  % Multisensory
    >{\RaggedRight\arraybackslash}X       % Application
}
\toprule
\rowcolor[rgb]{0.851,0.851,0.851}
\textbf{Location} & 
\textbf{Device} & 
\textbf{Principle \& Perception} & 
\textbf{Multisensory} & 
\textbf{Application} \\
\midrule
face & & & & Niijima et al. (\cite{niijima2016proposal}) \\
forehead & & Saito et al. (\cite{Saito_2021}) & & \\
mouth & Jingu et al. (\cite{jingu2023lipio}) & & & Jingu et al. (\cite{jingu2023lipio}) \\
tongue & Mukashev et al. (\cite{Mukashev_2023}) & Mukashev et al. (\cite{Mukashev_2023}) & & Mukashev et al. (\cite{Mukashev_2023}) \\
earlobe & & & Stanke et al. (\cite{Stanke_2023}) & Stanke et al. (\cite{Stanke_2023}) \\
neck & Kono et al. (\cite{Kono_2018}) & & & Tanaka et al. (\cite{Tanaka_2022}) \\
shoulder & & & & Seetohul et al. (\cite{Seetohul_2024}), Lopes et al. (\cite{Lopes_2017}) \\
upper arm & Dosen et al. (\cite{Dosen_2017}), Pfeiffer et al. (\cite{Pfeiffer_2016}) & Cheng et al. (\cite{cheng2024paired}), Blondin et al. (\cite{blondin2021perception}), Matsubara et al. (\cite{Matsubara_2023}), Kurita et al. (\cite{Kurita_2016}), Kim et al. (\cite{Kim_2022}), Dosen et al. (\cite{Dosen_2017}), Choi et al. (\cite{Choi_2016}), Ohara et al. (\cite{Ohara_2022}), Choi et al. (\cite{Choi_2017}) & & Elsharkawy et al. (\cite{elsharkawy2024sync}), Faltaous et al. (\cite{Faltaous_2022}), Lopes et al. (\cite{Lopes_2017}), Lopes et al. (\cite{Lopes_2015Affordance}), Lopes et al. (\cite{Lopes_2015Impacto}), Lopes et al. (\cite{Lopes_2018}) \\
elbow & & Cai et al. (\cite{Cai_2023}) & & \\
forearm & Stephens et al. (\cite{stephens2020comparison}), Shi et al. (\cite{Shi_2021}), Pfeiffer et al. (\cite{Pfeiffer_2016}), Abbass et al. (\cite{Abbass_2021}), Cheng et al. (\cite{Cheng_2017}), Kono et al. (\cite{Kono_2018}) & Akhtar et al. (\cite{akhtar2014relationship}), Djozic et al. (\cite{djozic2015psychophysical}), Dong et al. (\cite{Dong_2020}), Parsnejad et al. (\cite{Parsnejad_2019}), Shi et al. (\cite{Shi_2015}), Ishimaru et al. (\cite{Ishimaru_2022}), Tajima et al. (\cite{Tajima_2022}), Shahu et al. (\cite{Shahu_2022}), Ohara et al. (\cite{Ohara_2022}), Knibbe et al. (\cite{knibbe2018experiencing}), Franceschi et al. (\cite{Franceschi_2017}), Gholinezhad et al. (\cite{Gholinezhad_2022}), Yang et al. (\cite{Yang_2019}), Gehrke et al. (\cite{gehrke2019detecting}) & Gehrke et al. (\cite{gehrke2019detecting}), Korres et al. (\cite{korres2022comparison}), Dideriksen et al. (\cite{Dideriksen_2022electrotactile}), Pfeiffer et al. (\cite{Pfeiffer_2014}), Dr. Alonzo et al. (\cite{DrAlonzo_2014}) & Duente et al. (\cite{duente2018muscleio}), Elsharkawy et al. (\cite{elsharkawy2024sync}), Shindo et al. (\cite{Shindo_2021}), Pfeiffer et al. (\cite{Pfeiffer_2015Virtual}), Dideriksen et al. (\cite{Dideriksen_2022task}), Ishimaru et al. (\cite{Ishimaru_2020}), Lopes et al. (\cite{Lopes_2015Prop}), Lopes et al. (\cite{Lopes_2015Impacto}), Lopes et al. (\cite{Lopes_2015Affordance}), Faltaous et al. (\cite{Faltaous_2022}), Patibanda et al. (\cite{Patibanda_2023Fused}), Lopes et al. (\cite{Lopes_2016}), Lopes et al. (\cite{Lopes_2017}), Niijima et al. (\cite{Niijima_2023}), Zhou et al. (\cite{Zhou_2024}), Patibanda et al. (\cite{Patibanda_2024}), Patibanda et al. (\cite{Patibanda_2023Auto}), Lopes et al. (\cite{Lopes_2018}) \\
wrist & Tanaka et al. (\cite{Tanaka_2024}), Takahashi et al. (\cite{Takahashi_2024}) & Duente et al. (\cite{duente2023colorful}), Tanaka et al. (\cite{Tanaka_2023}), Tanaka et al. (\cite{Tanaka_2024}) & Lee et al. (\cite{Lee_2024}), Tanaka et al. (\cite{Tanaka_2024}), Stanke et al. (\cite{Stanke_2020}) & Tanaka et al. (\cite{Tanaka_2023}), Pohl et al. (\cite{Pohl_2018}), Lopes et al. (\cite{Lopes_2015Affordance}), Patibanda et al. (\cite{Patibanda_2023Auto}) \\
palm & Yao et al. (\cite{Yao_2022}), Huang et al. (\cite{Huang_2023}), Lin et al. (\cite{Lin_2024}) & Alotaibi et al. (\cite{alotaibi2022first}), Lin et al. (\cite{Lin_2022}), Yao et al. (\cite{Yao_2022}) & Huang et al. (\cite{Huang_2023}), Alotaibi et al. (\cite{Alotaibi_2020}) & Lin et al. (\cite{Lin_2022}), Pamungkas et al. (\cite{Pamungkas_2015}), Yao et al. (\cite{Yao_2022}) \\
back of hand & & Tanaka et al. (\cite{Tanaka_2023}) & Lee et al. (\cite{Lee_2024}) & Pamungkas et al. (\cite{pamungkas2019electro}), Lopes et al. (\cite{Lopes_2015Impacto}), Tanaka et al. (\cite{Tanaka_2023}) \\
finger & Groeger et al. (\cite{groeger2019tactlets}), Yao et al. (\cite{Yao_2022}), Garenfeld et al. (\cite{Garenfeld_2023}), Takami et al. (\cite{Takami_2023}) & An et al. (\cite{an2021tactile}), Yao et al. (\cite{Yao_2022}), Zhang et al. (\cite{Zhang_2022}), Yoshimoto et al. (\cite{Yoshimoto_2015}) & Yem et al. (\cite{Yem_2017}), Stanke et al. (\cite{Stanke_2020}) & Pamungkas et al. (\cite{pamungkas2019electro}), Takami et al. (\cite{Takami_2023}), Yao et al. (\cite{Yao_2022}) \\
fingertip & Lin et al. (\cite{Lin_2022}), Hummel et al. (\cite{Hummel_2016}), Isakovic et al. (\cite{Isakovic_2022}), Jingu et al. (\cite{jingu2023double}), Jingu et al. (\cite{Jingu_2024}), Withana et al. (\cite{Withana_2018}), Teng et al. (\cite{Teng_2024}), Lin et al. (\cite{Lin_2024}), Vizcay et al. (\cite{Vizcay_2022}), Zhou et al. (\cite{Zhou_2022Braille}) & Lin et al. (\cite{Lin_2022}), Cai et al. (\cite{Cai_2023}), Zhou et al. (\cite{Zhou_2022Explore}), Isakovic et al. (\cite{Isakovic_2022}), Kaczmarek et al. (\cite{Kaczmarek_2017}), Jingu et al. (\cite{Jingu_2024}), Rahimi et al. (\cite{Rahimi_2019Non}), Rahimi et al. (\cite{Rahimi_2019Adp}), Rahimi et al. (\cite{Rahimi_2019Dyn}) & Suga et al. (\cite{Suga_2024}), Suga et al. (\cite{Suga_2023}) & Lin et al. (\cite{Lin_2022}), Zhou et al. (\cite{Zhou_2022Braille}), Rahimi et al. (\cite{Rahimi_2022}), Trinitatova et al. (\cite{Trinitatova_2022}), Jiang et al. (\cite{Jiang_2024}), Kourtesis et al. (\cite{Kourtesis_2022}) \\
thumb & & & & Trinitatova et al. (\cite{Trinitatova_2022}), Parsnejad et al. (\cite{Parsnejad_2020Eva}), Yoshimoto et al. (\cite{Yoshimoto_2016}) \\
index finger & & Zhou et al. (\cite{Zhou_2023}) & Pfeiffer et al. (\cite{Pfeiffer_2015Virtual}) & Pfeiffer et al. (\cite{Pfeiffer_2015Virtual}), Parsnejad et al. (\cite{Parsnejad_2020Eva}), Yoshimoto et al. (\cite{Yoshimoto_2016}) \\
middle finger & & Zhou et al. (\cite{Zhou_2023}) & & \\
thigh & Pfeiffer et al. (\cite{Pfeiffer_2016}) & & & Hwang et al. (\cite{Hwang_2024}), Um et al. (\cite{Um_2024}), Pfeiffer et al. (\cite{Pfeiffer_2015Cruise}), Auda et al. (\cite{Auda_2019}) \\
lower leg & Pfeiffer et al. (\cite{Pfeiffer_2016}), Lopes et al. (\cite{Lopes_2015Impacto}) & & & Lopes et al. (\cite{Lopes_2015Impacto}), Hwang et al. (\cite{Hwang_2024}), Um et al. (\cite{Um_2024}) \\
ankle & & Zhang et al. (\cite{Zhang_2022}) & & \\
feet & Ushiyama et al. (\cite{Ushiyama_2023}) & Ushiyama et al. (\cite{Ushiyama_2023}) & Ushiyama et al. (\cite{Ushiyama_2023}) & \\
other & & Parsnejad et al. (\cite{Parsnejad_2020Use}), Parsnejad et al. (\cite{Parsnejad_2022}), Faltaous et al. (\cite{faltaous2024understanding}) & & \\
\bottomrule
\end{tabularx}
\end{table*}

The details of the publications are provided in \autoref{tab:papers}. The reviewed literature spans four main research thrusts: (1) electrical haptic feedback devices, (2) the underlying principles of electrical haptic stimulation and the relationship between stimulus and perception, (3) the integration and comparison of multimodal feedback, and (4) applied implementations in various domains of human-computer interaction. In particular, some publications address multiple aspects, leading to overlaps in categorization.

A stratified analysis of stimulation sites indicates a pronounced bias toward the upper limbs in current electrotactile research. Quantitative mapping reveals a disproportionate focus on five primary anatomical regions: the forearm (n=52, 47.3\%), fingertip (n=22, 20.0\%), upper arm (n=16, 14.5\%), finger (n=10, 9.1\%), and wrist (n=9, 8.2\%). Beyond these five regions, however, a considerable portion of the body remains largely unexplored. In light of the emerging metaverse vision, where users engage in fully immersive, networked experiences, haptic feedback should ideally encompass the entire body \cite{steinicke2016being}. Yet, given the current research landscape, there remains significant potential for further exploration in this domain.

In the following sections, we systematically discuss electrical haptic feedback across five key areas: Device (Section \ref{device}), Tactile Perception (Section \ref{tactile perception}), Kinesthetic Perception (Section \ref{kinesthetic perception}), Multisensory (Section \ref{multisensory}), and Application (Section \ref{application}), covering all relevant publications. Since some studies comprehensively address multiple aspects,such as device design, perception experiments, and applications, we examine them from multiple perspectives across different sections.

Electrical haptic feedback can be classified into tactile sensation and kinesthetic sensation. Tactile sensation refers to the perception of texture, shape, or the intensity of electrical stimulation, while kinesthetic sensation arises from muscle stimulation, leading to the perception of muscle contractions or movement-related sensations. Electrical stimulation can sometimes induce muscle contractions, resulting in the perception of kinesthetic sensations, or pseudo-kinesthetic sensations (perceived movement without actual motion). For consistency, we will collectively refer to these perceptions as kinesthetic sensations in the following content. Accordingly, we will analyze tactile and kinesthetic perception separately in Section \ref{tactile perception} and Section \ref{kinesthetic perception}.

Section \ref{device} focuses on device-related advancements in electrical haptic applications, including electrode design and wearable systems. Section \ref{multisensory} examines electrotactile feedback in comparison with other mainstream haptic feedback modalities, while Section \ref{application} explores its applications in human-computer interaction (HCI).

\subsection{Devices}
\label{device}
\begin{figure}[htbp]
    \centering
    \includegraphics[width=\linewidth]{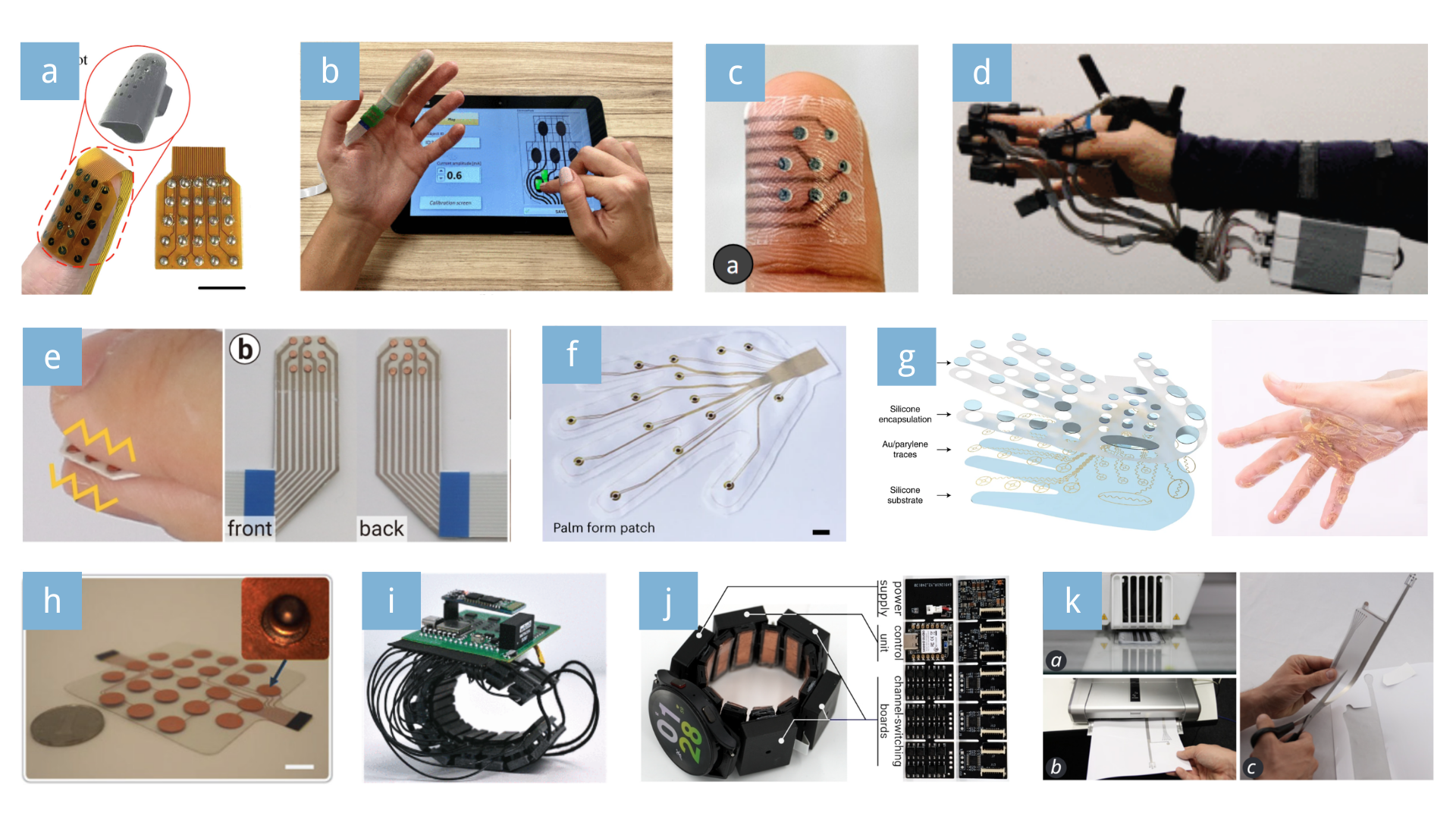}
    \caption{Electrotactile Devices. (a)–(e) Fingertip solutions from \cite{Lin_2022,Isakovic_2022,Withana_2018,Hummel_2016, jingu2023double}; (f)–(g) hand-worn systems proposed by \cite{Huang_2023,Yao_2022}; (h) a forearm interface from \cite{Shi_2021}; (i)–(j) wrist-based devices by \cite{Tanaka_2024,Takahashi_2024}; and (k) an example of using silver nanoparticle ink for electrode printing.}
    \label{fig:devices}
\end{figure}

Hardware-based electrotactile feedback devices fundamentally support electrotactile perception. Among the 110 reviewed articles, 29 examine such devices, encompassing research into both soft and rigid electrodes as well as soft and rigid device designs (\autoref{fig:devices}).

% electrode
Commonly used electrodes, such as metal electrodes and self-adhesive electrodes, suffer from several limitations, including bulkiness, limited resolution, unstable contact interface, and lengthy fabrication processes. Recent advances in electro-haptic electrode research have introduced innovative solutions. Compared to separated electrodes, concentric electrodes generate more localized sensations, reduce electromyographic (EMG) interference, and minimize the likelihood of painful or prickly side effects \cite{stephens2020comparison}. Additionally, emerging techniques such as conductive inkjet printing with silver-nanoparticle ink \cite{groeger2019tactlets, Garenfeld_2023} or conductive 3D printing using conductive PLA filament \cite{groeger2019tactlets} enable more convenient flex electrode production. Similarly, PEDOT:PSS has also been explored as a promising material for flex electrode fabrication \cite{Withana_2018, Garenfeld_2023}.

% fingertip device
The fingertip, governed by a high density of mechanoreceptors, is the primary medium for human touch \cite{cellis1977two}. However, its small size poses significant challenges in device design and fabrication. Traditional tactile devices are typically based on rigid PCBs and use metallic electrodes, such as copper, as interactive components \cite{Hummel_2016}. However, with advancements in flexible electronics, flexible devices have become mainstream. Flexible electrode arrays, such as those utilizing solder pads on flexible PCBs for fingertip haptic feedback \cite{Vizcay_2022, Zhou_2022Braille, Takami_2023} or arrays made of silver-coated copper spheres on flexible PCBs \cite{Lin_2022}, have demonstrated significant adaptability and high resolution. Further innovations, including Ag/AgCl electrodes \cite{Isakovic_2022} and compact arrays with nine or three electrodes \cite{jingu2023double, Jingu_2024}, enhance spatial perception at the fingertip. Material innovations, such as ultrathin fingertip interfaces combining PEDOT:PSS and Ag/AgCl electrodes \cite{Withana_2018, Garenfeld_2023} and perforated designs enabling direct skin contact with physical objects \cite{Teng_2024}, further advance the field. Additionally, high-resolution electrotactile interfaces seamlessly integrated into textiles \cite{Lin_2024} push the boundaries of wearable haptic technologies.

% hand and wrist device
In the design of haptic devices for the hand and wrist, the need to accommodate the stretching caused by hand movements has led to the growing adoption of flexible materials. For instance, hydrogel-based electrode patches \cite{Yao_2022}, flexible electronic arrays using stretchable silver ink and thermoplastic polyurethane films as conductive substrates \cite{Lin_2022}, or Cr–Au–PI thin films combined with soft silicone TC as a substrate are commonly employed to address these challenges. In contrast, wrist-mounted devices typically require less adaptation to complex curvatures, allowing flexibility to be achieved through either flexible structures or materials while incorporating rigid electrodes. For example, copper electrodes combined with conductive gel sheets and flexible structures or materials have been used to create wearable devices \cite{Takahashi_2024, Cheng_2017}. Similarly, high-biocompatibility elastomer electrodes integrated with bendable structures have been utilized to enhance flexibility \cite{Tanaka_2024}. Furthermore, studies have explored the use of carbon rubber electrodes and gold-plated electrodes for electrotactile stimulation on the wrist, demonstrating that gold-plated electrodes exhibit superior conductivity \cite{duente2023colorful}. 

% other location
In addition to fingertip, hand, and wrist-mounted devices, haptic technologies have also been explored for other body parts. For instance, the arm is a common site for feedback devices, often utilizing flexible PCBs and Ag/AgCl electrodes \cite{Dosen_2017, Abbass_2021}. A notable example is a TENG-based electrical stimulation system with a single electrode (using the skin as ground), which significantly simplifies electrode array designs \cite{Shi_2021}. Similarly, flexible PCBs and copper electrodes have been adapted for haptic devices targeting the mouth \cite{Mukashev_2023, jingu2023lipio} and the foot \cite{Ushiyama_2023}. Furthermore, due to the versatility of Electrical Muscle Stimulation (EMS) technology, specialized EMS toolkits have been developed to enable haptic feedback across multiple body sites, lowering technical barriers and expanding application potential \cite{Pfeiffer_2016, Kono_2018}. 

% conclusion
The development of haptic devices exhibits clear trends toward flexibility, high resolution, and adaptability across various body parts. While traditional rigid electrodes and devices have provided a foundational basis, they face limitations such as bulkiness, low resolution, and complex manufacturing processes. Recent advancements in materials, such as conductive inkjet printing, conductive 3D printing, and PEDOT:PSS, have enabled the creation of flexible, lightweight electrodes with enhanced performance. Flexible electronics have become mainstream, particularly for fingertip devices, where high spatial resolution and compact designs are critical. Innovations such as ultrathin interfaces, perforated designs, and flexible PCBs have further improved adaptability and user comfort. For hand and wrist applications, the integration of stretchable materials, such as hydrogel-based electrodes and elastomer structures, addresses the challenges of complex curvatures while maintaining functionality. Beyond the hand and wrist, haptic devices have extended to other body parts, including the arm, mouth, and foot, leveraging flexible PCBs, copper electrodes, and EMS technologies for versatile and simplified designs. However, limitations remain, as research continues to focus predominantly on hand-based tactile perception, with studies on other body parts still in relatively early stages. These trends and challenges collectively highlight the ongoing transition toward more efficient, user-friendly, and wearable electrotactile feedback systems.

\begin{figure*}[htbp]
    \centering
    \includegraphics[width=\linewidth]{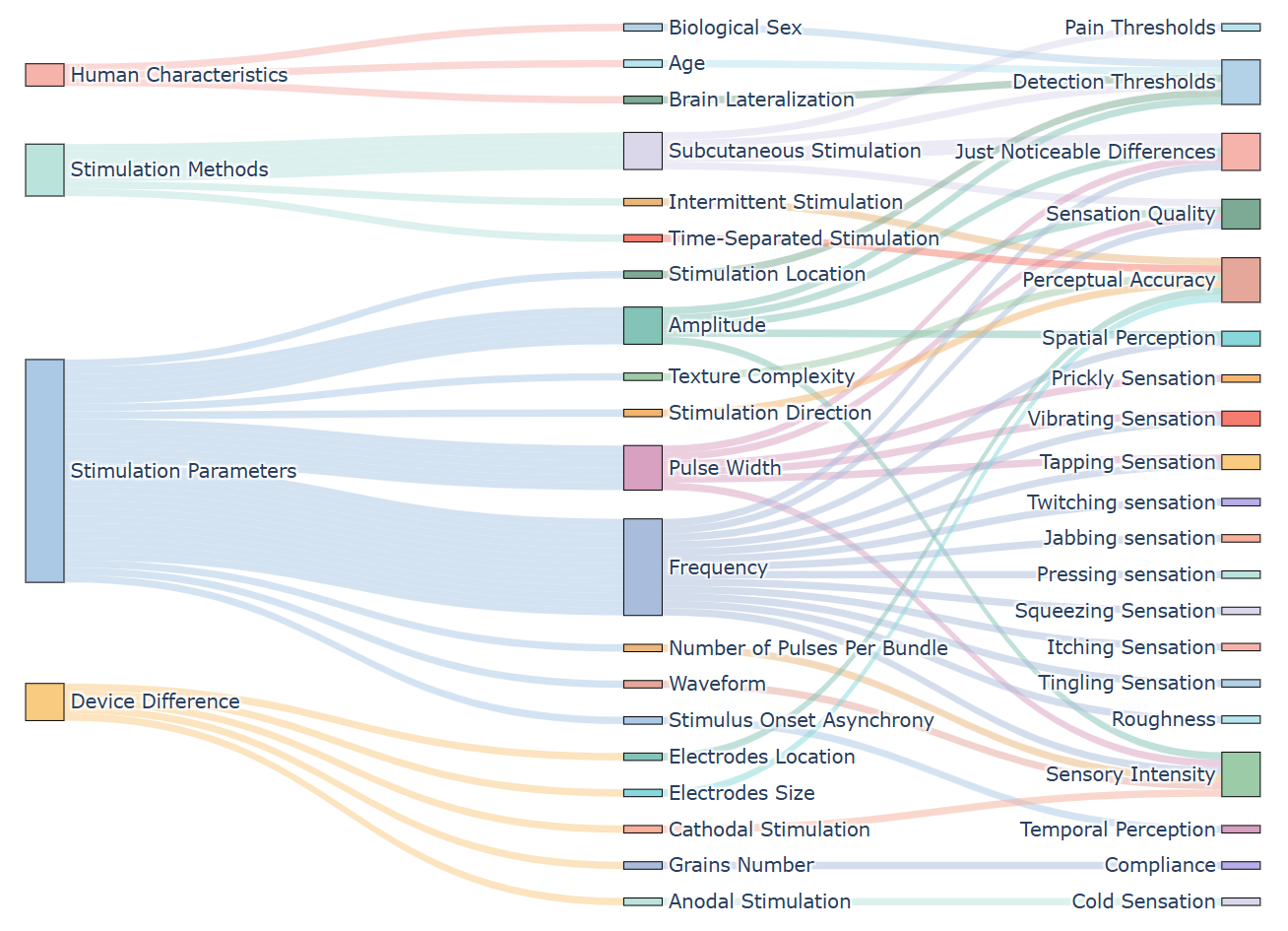}
    \caption{Flow Perception Diagram: This Sankey diagram visualizes the data flow from different classifications through influencing factors to relative sensations.}
    \label{fig:perception sankey}
\end{figure*}

\subsection{Tactile Perception}
\label{tactile perception}
In our investigation of tactile perception articles, we identified 49 experiments that explore various aspects of electrotactile stimulation. These studies encompass a range of topics, including electrical models of stimulation-skin impedance \cite{akhtar2014relationship, Rahimi_2019Non}, the impact of stimulation methods on perceptual ability \cite{Dong_2020, Choi_2016, alotaibi2022first, djozic2015psychophysical, Yao_2022, Cai_2023, Zhou_2022Explore, Yoshimoto_2015, Zhou_2023, Kaczmarek_2017, Parsnejad_2020Use, Parsnejad_2022, Parsnejad_2020Eva, Yang_2019}, and the relationship between stimulation parameters and affective load \cite{alotaibi2022first, Alotaibi_2020, blondin2021perception}. Additionally, they examine how electrotactile stimulation can induce diverse perceptual experiences, such as temperature perception \cite{Saito_2021}, roughness \cite{Lin_2022, Yoshimoto_2015}, softness \cite{Jingu_2024}, and multimodal sensations \cite{duente2023colorful, an2021tactile, Parsnejad_2019, Mukashev_2023}. Additionally, several studies have explored users' spatial perception \cite{Ushiyama_2023, blondin2021perception, Dosen_2017, Isakovic_2022, Tanaka_2023, Rahimi_2019Adp}, temporal perception \cite{Zhang_2022}, and spatiotemporal perception \cite{Lin_2022, Franceschi_2017, Gholinezhad_2022, Rahimi_2019Dyn} in response to electrotactile feedback.

We categorized and visualized the influencing factors and their corresponding perceptions (see \autoref{fig:perception sankey}). We found that device design, stimulation parameters, stimulation methods, and individual characteristics all impact tactile perception. These tactile perceptions include pain thresholds, detection thresholds, just noticeable differences, and perceptual accuracy. Among the stimulation parameters, frequency has been the most extensively studied in the current literature for its impact on tactile perception. Amplitude, on the other hand, primarily provides personalized perception parameters through individual calibration. The combination of pulse width and frequency enables various types of tactile sensations, such as roughness and pressure.

\begin{figure*}[htbp]
    \centering
    \includegraphics[width=\linewidth]{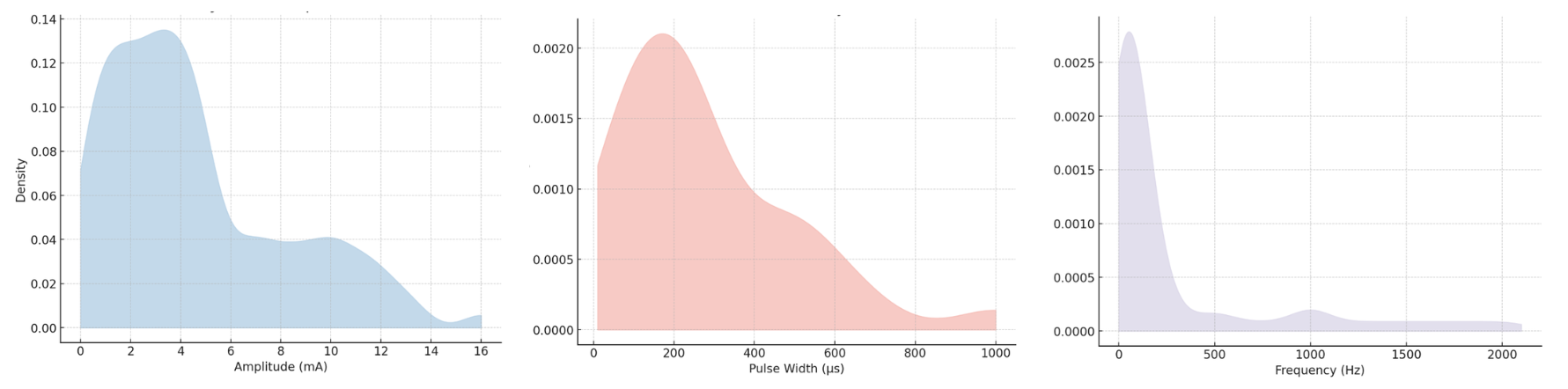}
    \caption{Density plots. The left displays the density plot of stimulation intensity, the middle represents the density plot of pulse width, and the right illustrates the density plot of stimulation frequency.}
    \label{fig:tactile_metrix}
\end{figure*}

In studies on tactile perception, square pulses are the predominant waveform used for electrotactile stimulation. However, the choice of electrode types is highly diverse, with no clear standardization across studies. The electrodes employed include gel electrodes, copper electrodes, self-adhesive electrodes, stainless steel electrodes, concentric electrodes, carbon rubber electrodes, silver-nanowire electrodes and Ag/AgCl electrodes. We extracted valid data from all studies on tactile perception and generated density plots for key electrical stimulation parameters, including stimulation intensity, pulse width, and frequency (see at \autoref{fig:tactile_metrix}). It is important to note that due to variations in data completeness across studies, only available and explicitly reported data were included in the density plots. Studies that did not provide relevant data were excluded from the analysis. Overall, the electrical stimulation parameters for tactile perception exhibit distinct concentration ranges. Stimulation intensity is primarily distributed between 0–6 mA, pulse width is concentrated within 0–400 µs, and stimulation frequency is mainly within the 0–300 Hz range.

\subsubsection{Electrical haptic models}

In the study of stimulation-skin impedance, researchers examined the relationships between peak resistance, pulse energy, and phase charge using gel electrodes, finding a strong linear correlation between peak impedance and both pulse energy and phase charge \cite{akhtar2014relationship}. They noted that $I^2 \times T$ (current squared × pulse duration) alone could not ensure consistent perception due to electrode-skin impedance variations, requiring compensation based on the linear relationship between phase charge and peak impedance. Similarly, the nonlinear properties of skin impedance were investigated, revealing significant impedance reductions over time, with current amplitude increasing up to fourfold within 15 minutes, though impedance recovered quickly. Real-time voltage adjustments using pulse-width modulation (PWM) and Kalman filtering were shown to effectively stabilize current and sensation \cite{Rahimi_2019Non}. Additionally, event-related potentials (ERPs), specifically Prediction Error Negativity (PEN), were introduced as an objective measure of haptic feedback, offering a novel way to quantify visuo-haptic mismatches in virtual reality and highlighting the need for adaptive stimulation control to address variations in skin impedance and sensory perception \cite{gehrke2019detecting}. 

Modeling electrotactile parameters enables quantitative predictions of subjective perception intensity (SI). Research establishes a linear relationship between parameter intensity (PI) and SI, allowing SI prediction under various electrode-skin conditions \cite{Zhou_2023}. Additionally, pulse amplitude (PA) and pulse width (PW) follow a logarithmic relationship, where increasing PA requires decreasing PW to maintain a constant perceived intensity. The parameter intensity metric $ PI = Log(PA) - 0.53 × Log(PW) $ effectively quantifies the perceptual impact of different stimulation parameter combinations \cite{Zhou_2023}.

\subsubsection{Perceptual ability}
Among all studies investigating perceptual ability, the research can be categorized into five main areas: the effect of stimulation methods on perception, differences in perception thresholds across body regions, the impact of frequency on perceptual accuracy, the influence of waveform and modulation on perception, and the effect of pulse width on perceptual sensitivity.

% the effect of stimulation methods on perception
In research on the effects of stimulation methods on perception, researchers investigated subcutaneous versus surface stimulation, finding that detection thresholds were more stable with subcutaneous stimulation, while pain threshold repeatability and just noticeable differences were better with surface stimulation. Psychophysical parameters, including dynamic range, resolution, sensory quality, intensity, and comfort, showed no significant variations across sessions for either type of stimulation, with subcutaneous stimulation offering better localization and fewer unintended sensory modalities \cite{Dong_2020}. Similarly, the effectiveness of dual-channel electrotactile stimulation in conveying tactile intensity information and improving perception accuracy was examined, revealing that intermittent stimulation achieved higher perception accuracy compared to continuous stimulation \cite{Choi_2016}.

% differences in perception thresholds across body regions
Studies on perceptual thresholds and sensitivity indicate that tactile perception varies by age, gender, and stimulation parameters. Research shows that males tend to have higher perception thresholds than females, and thresholds increase with age \cite{Yao_2022}. Fingertips exhibit the highest sensitivity (threshold < 1mA), while the palm center and knuckles are less sensitive (threshold > 1mA). Additionally, low-frequency stimulation (25 Hz) is perceived more easily than high-frequency stimulation (200 Hz). In the context of brain lateralization, dominant hands exhibit significantly higher electrotactile perception thresholds than non-dominant hands, although no significant differences were observed at the elbow \cite{Cai_2023}. Furthermore, pulse amplitude (PA) and pulse width (PW) significantly affect perceived intensity, with sensitivity varying across fingers, where the middle finger is the most sensitive, followed by the thumb and index finger \cite{Zhou_2022Explore}.

The ability to differentiate stimulation intensity and frequency has been explored across multiple studies. Users can clearly distinguish four intensity levels but struggle to differentiate frequencies above 30 Hz \cite{alotaibi2022first}. Perceptual resolution improves with higher frequency stimulation (10–100 Hz), and frequency is more influential than pulse width in sensory perception \cite{djozic2015psychophysical}. Low-frequency electrical stimulation (25 pps) enhances vibration perception, while high-frequency stimulation (100 pps) is associated with static pressure sensations rather than vibration \cite{Yoshimoto_2015}. Additionally, current amplitude significantly enhances frequency discrimination, whereas frequency variation has a more negligible effect on intensity perception \cite{Kaczmarek_2017}. Studies on high-frequency electrotactile stimulation (100 Hz–2.1 kHz) reveal that while such stimulation is detectable, it is difficult to differentiate specific frequencies \cite{Parsnejad_2020Use}. However, three distinct intensity levels can be perceived at package frequencies of 20 Hz and 40 Hz, with accuracy exceeding 85\% \cite{Parsnejad_2022}.

The role of waveform modulation in electrotactile perception has been extensively examined. Studies comparing square waves (CSW), sine waves (SW), and time-varying square waves (TPSW) reveal that TPSW provides the most dynamic tactile perception, simulating a rhythmic pressing sensation, while CSW is associated with the weakest and most uncomfortable tactile sensations \cite{Yang_2019}. Additionally, high-frequency pulse repetition rates (SRR > 1 kHz) do not interfere with bundle repetition rate (BRR) perception, and subjects can distinguish at least three BRR levels (18 Hz, 32 Hz, 67 Hz) \cite{Parsnejad_2020Eva}. Furthermore, modifying high-frequency pulse bundles (400 Hz) into low-frequency modulated signals (6 Hz) enables the creation of at least three distinct tactile sensations by adjusting the duty cycle \cite{Parsnejad_2020Use}. High package duty cycles (>80\%) enhance intensity perception but may obscure frequency discrimination \cite{Parsnejad_2022}.

\subsubsection{Affective load} Electrotactile stimulation has been investigated in the context of emotional burden and user perception, focusing on how stimulation parameters—including pulse width, amplitude, and frequency—influence urgency, annoyance, valence (pleasantness), and arousal.

Research has demonstrated that increasing stimulation intensity leads to higher urgency, annoyance, and arousal, but simultaneously decreases valence (pleasantness) \cite{alotaibi2022first}. Similarly, studies examining pulse width and frequency variations found that higher pulse widths and frequencies resulted in greater urgency and annoyance perception, though frequency had no significant impact on valence \cite{Alotaibi_2020}. Notably, stimulation frequencies above 30 Hz did not further affect perception, while higher pulse widths consistently reduced valence, albeit with no significant difference between high and low pulse widths.

User experience studies have also examined comfort levels associated with different electrotactile parameters. Research conducted on the forearm using a square wave pulse and a bracelet electrode array found that higher stimulation intensities (7 mA) were perceived as more painful and uncomfortable, while moderate intensities (3–5 mA) were considered more tolerable \cite{blondin2021perception}. Additionally, lower frequencies (35 Hz) were rated as more comfortable compared to higher frequencies (200 Hz), reinforcing the trend that high-frequency electrical stimulation tends to induce greater discomfort.

\subsubsection{Multimodal sensation} This section introduces the various perceptions induced by electrical stimulation, such as temperature perception, roughness perception, texture simulation, pressure perception, shape recognition and so on.

ES has been extensively studied for its ability to elicit a range of tactile perceptions by modulating electrical parameters. Research has shown that different electrode polarities influence temperature perception, with anodal stimulation more likely to induce cold sensations, while cathodal stimulation produces stronger pressure and vibration sensations \cite{Saito_2021}. Additionally, cathodal stimulation elicits stronger sensations than anodal stimulation \cite{Yem_2017}.

The frequency and pulse width of electrotactile stimulation play a crucial role in determining the type of tactile sensations experienced. Low-frequency stimulation (< 50 Hz) typically induces twitching and pricking sensations, while medium-frequency stimulation (50–250 Hz) elicits vibrating and stimulating perceptions. In contrast, high-frequency stimulation (> 250 Hz) is associated with squeezing and itching sensations \cite{duente2023colorful}. Furthermore, modulating the frequency of electrical stimulation within a single electrode can generate two distinct tactile sensations: pressure and tapping \cite{Choi_2017}. These findings suggest that frequency alone can be utilized to create multiple tactile perceptions without the need for additional electrodes or channels, offering a compact and efficient approach to designing electrotactile interfaces. Recognition accuracy varied across frequency and pulse width combinations, with prodding sensation being most recognizable (84\%  at 2 Hz, 10 µs) and squeezing sensation the least (14\% at 1000 Hz, 80 µs), highlighting the critical role of frequency modulation in optimizing tactile feedback \cite{duente2023colorful}.

Apart from periodic stimulation, aperiodic ES patterns have demonstrated the capability to generate diverse tactile experiences such as softness, diffusion, pressing, stretching, and tapping \cite{an2021tactile}. Moreover, waveform parameters significantly affect perception, with specific waveforms being associated with distinct sensations, including vibration, prickling, and tapping. Notably, subjects achieved a 94\% accuracy in distinguishing different waveforms, highlighting the potential for fine-tuned haptic feedback design \cite{Parsnejad_2019}. In line with this, a wearable wrist feedback device designed by \cite{Tanaka_2024} elicited commonly perceived sensations such as "tapping," "pressing," and "tingling," further supporting the haptic sensation. Beyond general tactile sensations, ES applied to the tongue reveals regional differences in perception. The tongue tip is the most sensitive, capable of detecting prickling, softness, and salty taste, while the tongue root shows weaker responses. High-frequency stimulation (150 Hz) is more effective in inducing taste perceptions than low-frequency stimulation (100 Hz), with salty taste being the most frequently reported, followed by sour taste. Additionally, higher pulse widths (300 µs) increase the likelihood of prickly sensations \cite{Mukashev_2023}.

ES has also been explored for simulating surface textures, with different stimulation parameters enabling the perception of various materials. Low-frequency, high-voltage stimulation effectively mimics rough textures, such as rock surfaces, while high-frequency, low-voltage stimulation replicates smooth textures like glass. Recognition accuracy for texture differentiation has been reported at 98\%, demonstrating the fidelity of ES-based texture simulation \cite{Lin_2022}. Furthermore, low-frequency stimulation (20–60 pps) enhances roughness perception, whereas high-frequency stimulation (80–100 pps) reduces tactile distinctiveness due to sensory blending \cite{Yoshimoto_2015}.

In addition to texture perception, ES influences compliance perception and shape recognition. Different numbers of activated electrodes generate distinct compliance sensations, with a minimum of four electrodes being sufficient for perception and nine electrodes producing the most pronounced compliance effects \cite{Jingu_2024}. Shape perception through ES has been validated, with subjects accurately perceiving and sketching contour shapes based on electrode activation patterns. Larger shapes, such as squares, are more easily recognized, and perception accuracy is slightly lower in the vertical direction compared to the horizontal direction \cite{Jingu_2024}. 

\subsubsection{Spatial, temporal, and spatiotemporal perception} 

Studies on spatial perception show that electrode placement, stimulation frequency, and encoding methods significantly impact localization accuracy. On the arm, studies show that frequency variation proves more effective than intensity changes \cite{blondin2021perception}. In contrast, a mixed-frequency encoding strategy improves grip force perception accuracy, surpassing purely spatial encoding methods \cite{Dosen_2017}. On the hand and fingertip, the size and placement of electrodes also affected localization accuracy, with smaller reference electrodes shifting perception areas and larger ones maintaining accuracy \cite{Isakovic_2022}. Temporal perception studies revealed that users could distinguish two consecutive stimuli reliably when Stimulus Onset Asynchrony exceeded 60 ms, but accuracy declined for intervals below this threshold \cite{Zhang_2022}.  Shape recognition on the foot achieved a 47\% accuracy, outperforming vibration feedback \cite{Ushiyama_2023}, while two-point discrimination on the foot was 29 mm on average \cite{Ushiyama_2023}. 

Spatiotemporal perception studies further investigated dynamic electrotactile stimuli, focusing on shape recognition, motion patterns, and frequency discrimination. High-resolution pattern recognition on the palm and fingertip achieved 96\% accuracy, with a spatial resolution of $76 dots/cm²$ \cite{Lin_2022}. Shape identification on the forearm showed recognition rates of 86\% for lines, 73\% for geometric figures, and 72\% for letters, with path-tracking accuracy reaching 98\% \cite{Franceschi_2017}. Additionally, studies on frequency-based discrimination found that time-separated stimuli provided better differentiation than continuous stimuli \cite{Gholinezhad_2022}. Motion perception experiments demonstrated that vertical motion patterns were easier to recognize than horizontal ones, likely due to natural finger curvature affecting electrode contact \cite{Rahimi_2019Dyn}. These findings contribute to the development of more precise and dynamic electrotactile feedback systems for applications in virtual reality, assistive haptics, and prosthetic technologies.

\subsection{kinesthetic Perception}
\label{kinesthetic perception}
\begin{figure*}[htbp]
    \centering
    \includegraphics[width=\linewidth]{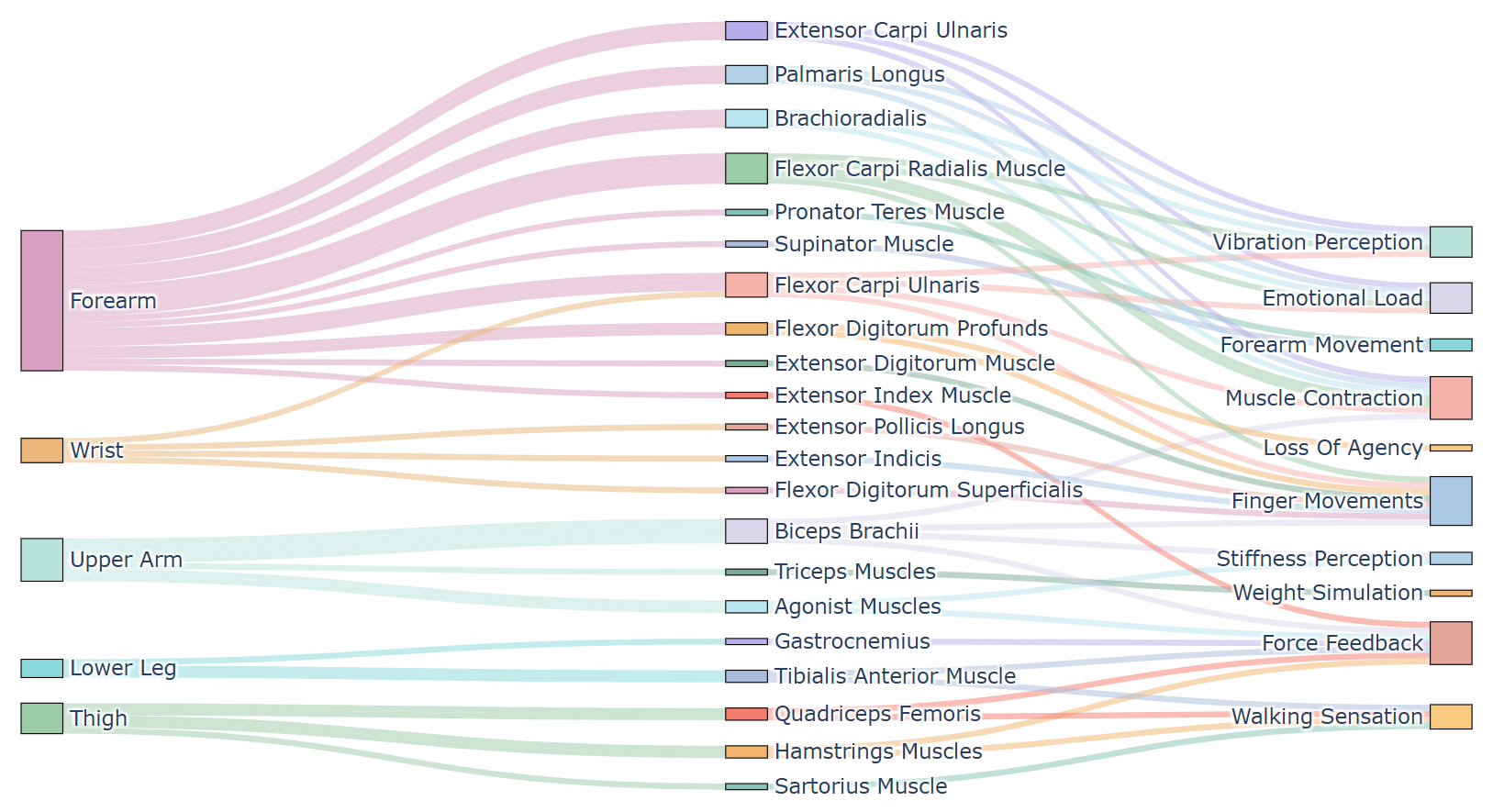}
    \caption{Flow Perception Diagram: This Sankey diagram visualizes the data flow from stimulation location through stimulation muscles to relative sensations.}
    \label{fig:kinestheticSankey}
\end{figure*}

A total of 19 studies have investigated kinesthetic perception induced by electrical muscle stimulation (EMS). Advancements in EMS-based haptics have demonstrated significant improvements in haptic perception, force feedback simulation, motion assistance, and VR locomotion. While EMS enables precise muscle stimulation for enhanced interaction, challenges such as discomfort, agency control, and signal optimization remain critical areas for future research. Research on kinesthetic perception has predominantly focused on the arm (see \autoref{fig:kinestheticSankey}), as it contains a large number of muscles that influence both arm and hand movements. This aligns with the broader research trends in tactile perception. Notably, the wrist serves as a natural stimulation site due to the passage of muscle groups through this region. Electrical stimulation applied to the arm can elicit various sensations, including vibrations, muscle contractions, arm and hand movements, force feedback, and the perception of rigid objects. Additionally, some studies have explored tactile sensations induced by leg stimulation, with primary attention given to force feedback and walking sensations.

\subsubsection{Haptic perception and muscle response}
Studies on EMS-based sensory perception have demonstrated its capability to enhance haptic feedback and adapt to individual differences. A system utilizing a millimeter-scale electrode array successfully identified and stimulated precise motor points (MPs), which were found to shift based on elbow angle and hand orientation, while smaller electrodes resulted in higher muscle contraction intensity \cite{Matsubara_2023}. To improve the practicality of wearable EMS applications, a wrist-based EMS control system was introduced, reducing calibration time by 50\% compared to traditional forearm EMS systems, while being perceived as more socially acceptable for daily use \cite{Takahashi_2024}. Research optimizing electrical stimulation for cutaneous haptic feedback established that pulse width (from 0.2 to 5 ms), frequency (from 45 to 70 Hz), and intensity (less than or equal to 0.5 mA) provided clear but painless sensations, with deeper sensations induced by larger pulse widths and higher frequencies \cite{Shi_2015}.

Another study investigated the influence of EMS waveform on discomfort and force output, revealing that pulse waveforms minimized discomfort while maintaining stable force, while reverse sawtooth waves reduced perceived pain but decreased force output \cite{Ishimaru_2022}. The concept of agency in EMS-driven interactions was also explored, showing that when EMS actions aligned with user expectations, users attributed the action to themselves, whereas unexpected outcomes led to external attribution of control \cite{Tajima_2022}. Additionally, a dual-channel EMS system (DualEMS) was proposed for interferential current stimulation, effectively activating deep forearm muscles for pronation and supination, with high-frequency waves (from 4000 to 4080 Hz) reducing skin discomfort \cite{Ohara_2022}. Lastly, the relationship between EMS parameters and sensory perception was analyzed, revealing that low-frequency stimulation (from 20 to 55 Hz) produced rhythmic sensations, while higher frequencies (from 90 to 120 Hz) generated continuous perceptions like vibration and pressure \cite{knibbe2018experiencing}.

\subsubsection{Simulating external forces}
Electrical muscle stimulation (EMS) has been widely explored as a method for generating external force sensations, enhancing user interactions across various applications. One notable technique, Paired-EMS, stimulates agonist and antagonist muscles simultaneously, which has been shown to improve the realism, coordination, and enjoyment of force feedback \cite{cheng2024paired}. Research on force perception using EMS in muscles such as the quadriceps, hamstrings, tibialis anterior, and gastrocnemius revealed that users are more sensitive to horizontal force variations compared to vertical ones \cite{Hwang_2024}. Additionally, EMS-induced muscle contractions have been utilized to simulate stiffness, with findings showing a strong correlation between EMS intensity and the perceived stiffness \cite{Kurita_2016}. Additionally, EMS has been combined with visual distortion to expand the perceivable weight range by 2.5 times compared to visual feedback alone, significantly enhancing the perception of weight \cite{Kim_2022}. Building on these advancements, \cite{Lopes_2017} achieved multimodal “walls” in virtual worlds by coupling EMS with other sensory cues to replicate solidity and weight. Further efforts by Lopes et al.\cite{Lopes_2018} extended this concept, channeling virtual objects’ friction, gravity, and collisions via force feedback while allowing users to retain full use of their hands in the real world.

\subsubsection{Virtual locomotion and gait simulation}
EMS has been integrated into VR locomotion systems to enhance walking sensations in seated users. A system called LegSense induced gait-related muscle activation to simulate walking sensations without actual movement, significantly improving immersion and body ownership compared to visual-only cues \cite{Um_2024}. Another approach combined EMS with vision shift techniques to enable infinite walking within limited physical spaces, achieving the smallest physical walking radius (5.48m) when using EMS + vision shift, compared to 8.70m (EMS only) and 6.37m (vision shift only) \cite{Auda_2019}.

\subsection{Multisensory}
\label{multisensory}
A total of 17 articles compared electrical feedback with other feedback modalities, including vibrotactile feedback, force feedback, and hybrid multimodal feedback. This section categorizes and summarizes the findings from recent research on these modalities.

\subsubsection{Electrotactile feedback vs. vibrotactile feedback} Electrotactile and vibrotactile feedback have been widely compared across various applications, with each modality offering unique advantages depending on the context.

On the forearm, both electrotactile and vibrotactile feedback significantly improve response times compared to visual feedback. However, electrical stimulation generally achieves lower error rates, making it more suitable for high-precision tasks \cite{korres2022comparison}. Psychophysical studies further highlight that electrical stimulation provides higher spatial resolution and lower Just Noticeable Difference (JND) values than vibrotactile feedback, offering more precise tactile sensations. Interestingly, when both modalities are optimized for their respective best frequency ranges (100 Hz for electrical stimulation and 200 Hz for vibrotactile feedback), their performance becomes comparable \cite{Dideriksen_2022electrotactile}. In some applications, such as tactile augmented reality, electrical stimulation has demonstrated greater effectiveness in simulating larger virtual bumps, outperforming vibrotactile feedback in terms of realism and user perception. Additionally, compared to shear force feedback, electrical stimulation produces similar effects but offers a more compact and simplified setup \cite{Ishimaru_2020}. For 3D spatial tasks, electrical haptic also slightly outperforms vibrotactile feedback by providing more precise mappings to target locations \cite{Pfeiffer_2015Cruise}.

On the wrist, electrotactile has shown great promise in mixed-reality interactions. A wrist-worn electrical haptic device has been found to enhance remote tactile perception, often leading to a user preference over traditional vibrotactile wristbands \cite{Tanaka_2024}. Furthermore, studies comparing wrist- and ring-based tactile notifications reveal that electrical haptic feedback is more spatially localized, allowing for greater precision. In contrast, vibrotactile feedback tends to spread across a larger surface area, resulting in less precision but greater comfort for users \cite{Stanke_2020}.

For finger-based applications, electrotactile and vibrotactile feedback exhibit distinct characteristics depending on stimulation intensity. At lower intensities, electrotactile closely mimics mechanical stimulation, making it difficult for users to distinguish between the two. At higher intensities, however, anodic electrical stimulation produces sharper and more localized sensations, while vibrotactile feedback delivers more diffused sensations \cite{Yem_2017}.

\subsubsection{Electrotactile feedback vs. other modalities}
Beyond vibrotactile feedback, electrotactile has been compared to other modalities, including force feedback and visual cues. In material interaction tasks, electrotactile was found to be superior for simulating interactions with hard objects, while both electrotactile and vibrotactile feedback performed similarly for soft object interactions \cite{Pfeiffer_2014}. Additionally, in foot-based haptic systems, electrotactile enables higher spatial resolution, allowing users to simultaneously perceive both virtual and real-world terrain more effectively than vibrotactile feedback \cite{Ushiyama_2023}. \cite{Stanke_2023} conducted a comparative study on nine different notification modalities applied to the earlobe, including vibrotactile, electrotactile, pinprick, thermal feedback (cold/warm), private audio, LED, display, and public audio cues. The study evaluated these modalities in terms of reaction time, error rate, and user acceptability for both private and public notifications. The findings indicate that electrotactile stimulation is the most suitable for private notifications, as it yielded the fastest reaction time among all tested modalities.

\subsubsection{Hybrid feedback approaches}
Multimodal feedback has been investigated to improve user immersion and perception in virtual environments. Studies measuring event-related potentials (ERP) indicate that multimodal feedback enhances immersion, with electrotactile providing the strongest sense of realism despite vibrotactile feedback being perceived as more comfortable \cite{gehrke2019detecting}. Similarly, in VR-based Fitts’ Law tasks, electrotactile feedback significantly improves task accuracy, whereas vibrotactile feedback exhibits the weakest performance in both accuracy and reaction time \cite{Kourtesis_2022}.

Hybrid feedback systems integrating electrotactile with other modalities have been shown to enhance haptic perception and interaction accuracy. The Hybrid Vibro-Electrotactile (HyVE) interface significantly improves stimulus recognition accuracy, achieving a 98\% success rate, which is higher than that of vibrotactile-only interfaces and comparable to pure electrotactile feedback \cite{DrAlonzo_2014}.

In shape recognition tasks, combining force feedback with electrotactile improves accuracy and efficiency, as electrotactile provides additional edge perception cues, which enhance spatial awareness \cite{Suga_2023, Suga_2024}. Moreover, integrating electrotactile with vibrotactile feedback in virtual collision scenarios enhances realism and expressiveness. However, electrotactile-induced tingling sensations may be perceived as either beneficial or disruptive, depending on the application (e.g., enhancing experience in boxing but being undesirable in tennis) \cite{Lee_2024}.

Additionally, incorporating temperature feedback with electrotactile and vibrotactile stimulation results in the highest tactile recognition accuracy (>85\%), whereas unimodal visual feedback performs the worst (<50\%) \cite{Huang_2023}. These findings suggest that multimodal feedback systems, particularly those combining electrotactile with other tactile stimuli, offer significant advantages in haptic perception and user experience.

\subsubsection{Summary of multisensory comparison} Electrical feedback offers notable advantages over vibrotactile and other haptic modalities, including higher spatial resolution, lower error rates, and a more precise perception of tactile stimuli. However, vibrotactile feedback remains widely used due to its intuitive and comfortable nature. The choice between these modalities depends on the specific application, with electrotactile being preferable for high-precision tasks and vibrotactile feedback being better suited for general consumer applications.

Multimodal feedback integration has been shown to improve realism, accuracy, and user experience, with hybrid solutions combining electrotactile, vibrotactile, force, and temperature feedback demonstrating superior performance across multiple domains. Future research should explore optimizing these hybrid approaches to maximize the effectiveness of haptic feedback in diverse applications, ranging from medical simulations to immersive virtual reality environments.

\subsection{Applications}
\label{application}
Traditionally, researchers have relied on commercially available electrodes, such as gel-based or self-adhesive designs, to stimulate muscles and deliver haptic feedback. While these conventional solutions tend to be bulkier and offer lower spatial resolution, they have demonstrated substantial potential across a wide range of applications. For instance, electrotactile feedback has been shown to accelerate user learning and skill acquisition. Additionally, it has proven effective in enhancing notifications, facilitating navigation, and dynamically modulating cognitive load. In more unconventional implementations, electrical stimulation has been employed to alter perceived body ownership and introduce asymmetric interactions, thereby increasing immersion in interactive systems. Furthermore, electrotactile cues play a crucial role in VR and AR environments, enabling more immersive haptic feedback, realistic object interactions, and specialized training scenarios. As electrotactile technology continues to evolve, these applications underscore its versatility and growing influence in redefining human-computer interaction paradigms.

\subsubsection{Learning and training}
A key research focus in electrotactile haptics is its role in improving learning outcomes. In the domain of braille recognition, Lin et al. \cite{Lin_2022} introduced a high-resolution fingertip interface tailored for blind users, while Shi et al. \cite{Shi_2021} explored dynamic braille displays that enabled users to write characters based on tactile feedback. Further advancements by Zhou et al. \cite{Zhou_2022Braille} and Rahimi et al. \cite{Rahimi_2022} leveraged both flexible and rigid interface designs to enhance reading experiences for visually impaired individuals. Beyond textual applications, Faltaous et al. \cite{Faltaous_2022} demonstrated that EMS-based sign language training significantly enhanced user experience, learning efficiency, and memory retention. Similarly, Pamungkas et al. \cite{pamungkas2019electro} found that incorporating TENS-driven electrotactile feedback into finger movement training resulted in superior learning outcomes compared to purely audiovisual feedback. In the context of data interpretation, Jiang et al. \cite{Jiang_2024} showed that finger-edge electrotactile cues facilitated the comprehension of common graph types, while Zhou et al. \cite{Zhou_2024} proposed EMS-driven hand–eye coordination training, leading to substantially improved attention-switching speeds and long-term performance retention over a week.

Electrotactile feedback has also demonstrated its potential in skill-based training. Yoshimoto et al. \cite{Yoshimoto_2016} developed a specialized training system for dental wax carving, where electrotactile cues provided real-time feedback, helping users refine their carving techniques, reduce mistakes, and enhance sculpting accuracy. Such applications illustrate the broader implications of electrotactile stimulation beyond traditional learning environments, extending its benefits to precision-based skill development and professional training. Additionally, Shindo et al. \cite{Shindo_2021} investigated the use of fingertip electrical stimulation to stabilize user posture, demonstrating that EMS-assisted gentle contact effectively reduced sway compared to no-contact conditions, highlighting its potential in enhancing balance control and learning-related tasks.

\subsubsection{Teleoperation and advanced interaction}
Electrotactile feedback has been extensively explored in teleoperated and interactive systems. Trinitatova et al. \cite{Trinitatova_2022} combined electrotactile stimulation with kinesthetic feedback for remote manipulation, thereby conveying contact surfaces and pressure distributions. Similarly, Dideriksen et al. \cite{Dideriksen_2022task} investigated remote closed-loop control leveraging electrical cues, and Pamungkas et al. \cite{Pamungkas_2015} emphasized how electrotactile signals can heighten immersion in teleoperated settings. Outside teleoperation, electrotactile technology has also fueled novel device designs. For example, Lopes et al. \cite{Lopes_2015Prop, Lopes_2015Affordance} discovered that electrical muscle stimulation (EMS) could endow objects with dynamic affordances, effectively “transferring” usage instructions to the user. Meanwhile, Tanaka et al. \cite{Tanaka_2022} proposed a head orientation system that rotates the user’s head around yaw and pitch axes via EMS. Additional work, such as Lopes et al. \cite{Lopes_2016}, combined muscle stimulation and pen-based input to enable complex curve drawing and spatial output. 

Beyond entertainment, Mukashev et al. \cite{Mukashev_2023} introduced a novel tongue-based electrotactile interface, TactTongue, which frees users' hands and offers a new approach to VR interactions. Compared to traditional Arduino-based implementations, TactTongue significantly enhances the efficiency of haptic feedback development while reducing the requirement for programming expertise. User evaluations, collected through surveys, highlighted its intuitiveness, ease of use, and effectiveness, demonstrating its potential for expanding interaction possibilities in immersive virtual environments. Similarly, Jingu et al. \cite{jingu2023lipio} showed that even lips could serve dual roles as an input and output surface, freeing the user’s hands for other interactions.

\subsubsection{Notifications and navigation}
Electrical haptic stimuli also function as an effective communication channel for notifying users about events or navigational cues. Duente et al. \cite{duente2018muscleio} found that varying EMS intensity conveys urgency, even at the cost of lower comfort. In more private or precise scenarios, Stanke et al. \cite{Stanke_2023} introduced ear-lobe stimulation for confidential alerts, and Seetohul et al. \cite{Seetohul_2024} designed a shoulder-mounted device to guide users in 3D space. Pohl et al. \cite{Pohl_2018}, in contrast, documented how induced itching via EMS might be more suitable for emergency or negative alerts. Relatedly, Pfeiffer et al. \cite{Pfeiffer_2015Cruise} explored EMS-driven pedestrian navigation by stimulating leg muscles, highlighting a reduced dependence on visual or auditory cues.

\subsubsection{Body ownership and asymmetric interactions}
Another compelling trend is the adoption of “body lending,” where users partially surrender bodily control to a computer. Lopes et al. \cite{Lopes_2015Prop} discovered that muscle stimulation can produce a sense of blurred ownership, enhancing gameplay with “non-symmetric” interactions. Building on this concept, Patibanda et al. \cite{Patibanda_2023Fused} demonstrated “Fused Spectatorship,” where audience members “loan” their bodies to a computer through EMS to engage with the game, and Patibanda et al. \cite{Patibanda_2024} further advanced this notion under “Shared Bodily Fusion,” allowing multiple users to share control and foster richer inter-body interactions. Such designs remove barriers between the physical body and virtual environments, improving playability. 

\subsubsection{Other Immersive VR/AR haptics application}
Elsharkawy et al.\cite{elsharkawy2024sync} demonstrated that TENS-integrated electrotactile cues enhance user presence, reduce workload and heart rate, and alleviate motion sickness. Meanwhile, Pfeiffer et al. \cite{Pfeiffer_2015Virtual} leveraged electrotactile signals to improve object selection in 3D environments. Researchers have also leveraged EMS to enhance immersion and realism in mixed reality (MR) or VR applications. Lopes et al. \cite{Lopes_2017} achieved multimodal “walls” in virtual worlds, coupling EMS with other cues to replicate solidity and weight. Further efforts—Lopes et al. \cite{Lopes_2018}—channeled virtual objects’ friction, gravity, and collisions via force feedback while preserving the user’s hands for real-world interactions. Additionally, Niijima et al. \cite{niijima2016proposal} presented a method for rendering food textures by stimulating the facial muscles. By modulating EMS intensity and stimulation duration, the system allows users to perceive variations in hardness and elasticity, further demonstrating EMS’s versatility in sensory manipulation.

In VR locomotion, Hwang et al. \cite{Hwang_2024} proposed ErgoPulse to deliver scaled, biomechanically informed lower-limb feedback, and Um et al. \cite{Um_2024} elicited walking sensations for seated VR users without actual leg motion. Auda et al. \cite{Auda_2019} combined EMS with vision shifting techniques to simulate infinite walking in constrained physical spaces. 

Meanwhile, smartphone-based experiences have also benefited: Takami et al. \cite{Takami_2023} introduced ExtEdge, which augments a phone’s edges with EMS-based tactile content, and Tanaka et al. \cite{Tanaka_2023} created a full-palm haptic technique enabling distal tactile perception on the palm side. Yao et al. \cite{Yao_2022} designed an AR game in which a virtual mouse moves dynamically across the user's hand, leaping from one finger to another before settling at the base of the palm. Users perceive the movement and landing of the virtual mouse through electrotactile feedback, enhancing the sense of immersion and realism in the interaction.

\section{Discussion}
\subsection{Summary of Key Findings}
\subsubsection{Device}
The evolution of haptic devices demonstrates significant progress in material innovation and design flexibility, addressing the limitations of traditional rigid electrodes and devices. While early designs relied heavily on copper or gel-based electrodes, their bulkiness, limited precision, and lengthy fabrication processes have driven the exploration of advanced materials and fabrication techniques. Recent innovations, such as concentric electrodes, have improved sensory localization and reduced EMG interference, while emerging manufacturing methods, including conductive inkjet printing and 3D printing, have streamlined production processes \cite{groeger2019tactlets, Garenfeld_2023}. PEDOT:PSS has also emerged as a promising material, enabling the creation of lightweight, flexible electrodes \cite{Withana_2018, Garenfeld_2023} and flexible PCBs and ultrathin interfaces enhancing adaptability and user comfort \cite{Vizcay_2022, Zhou_2022Braille}. Similarly, hand and wrist devices have incorporated stretchable materials, such as hydrogel-based electrodes and elastomer structures, to address the challenges posed by complex curvatures \cite{Yao_2022, Tanaka_2024}. Beyond these applications, haptic technologies have been extended to other body parts, including the arm, mouth, and foot, leveraging flexible PCBs and EMS technologies to enable versatile and simplified designs \cite{Dosen_2017, Shi_2021, Mukashev_2023}. However, despite these advancements, research remains heavily focused on hand-centric devices, with applications for other body parts still in their infancy. Future efforts should expand the scope of research to develop comprehensive, user-friendly haptic systems applicable across diverse body sites.

\subsubsection{Perception}
Electrotactile stimulation has demonstrated a wide range of tactile perceptual capabilities, encompassing fundamental mechanisms such as stimulation-skin impedance modeling, perceptual thresholds, waveform modulation effects, and multimodal sensory integration. Our review of tactile perception studies highlights the predominant role of square pulse stimulation, with considerable variability in electrode materials and configurations, reflecting the lack of standardization in electrotactile research. The findings reveal distinct parameter ranges for effective electrotactile perception, with stimulation intensities typically between 0–6 mA, pulse widths within 0–400 µs, and frequencies concentrated around 0–300 Hz.

A key challenge in electrotactile perception is the significant variability in skin impedance, which affects current flow and perceived intensity. Dynamic compensation mechanisms, such as real-time pulse-width modulation and impedance monitoring, have been proposed to stabilize perception \cite{Rahimi_2019Non}. Additionally, electrotactile stimulation has been successfully linked to event-related potentials (ERPs), opening new avenues for objective measurement of haptic perception rather than relying solely on subjective reports \cite{gehrke2019detecting}.

Electrotactile stimulation further enables rich and diverse sensory experiences, including roughness, softness, and temperature perception, by modulating waveform and frequency parameters \cite{Saito_2021, Lin_2022, Jingu_2024}. Multimodal integration studies have shown that hybrid stimulation combining electrotactile cues with vibrotactile or visual feedback enhances realism and user engagement \cite{duente2023colorful, an2021tactile}. Furthermore, spatial and temporal perception studies confirm that electrotactile feedback can accurately convey motion, shape, and dynamic stimuli, with users achieving high localization accuracy under optimized stimulation conditions \cite{Ushiyama_2023, Zhang_2022}.

Although various sensory simulations have been successfully demonstrated, the haptic perception induced by electrical stimulation remains in its early stages. In the following discussion (Section \ref{sec:limitations}), we examine its limitations in greater detail.

\subsubsection{Multisensory}
Electrotactile stimulation has been widely compared with vibrotactile and other haptic modalities, demonstrating higher spatial resolution and lower error rates, making it more suitable for high-precision applications such as spatial mapping \cite{korres2022comparison, Pfeiffer_2015Virtual}. While electrotactile stimulation provides sharper and more localized sensations, vibrotactile feedback remains preferred for general applications due to its comfort and familiarity \cite{Yem_2017, Stanke_2020}. Additionally, electrotactile stimulation has been found to outperform vibrotactile feedback in augmented reality applications by providing more realistic virtual surface textures and better spatial perception \cite{Ishimaru_2020}.

Multimodal integration has further enhanced electrotactile perception by combining electrotactile with vibrotactile, force, and thermal feedback. Hybrid vibro-electrotactile systems (HyVE) achieve superior recognition accuracy, and electrotactile-force feedback integration improves shape perception \cite{DrAlonzo_2014, Suga_2023}. While multimodal approaches enhance realism and accuracy, balancing sensory realism and user comfort remains a challenge, as electrotactile-induced sensations may be perceived as tingling or intrusive in certain applications \cite{Lee_2024}.

Future research should focus on optimizing multimodal feedback strategies, refining adaptive stimulation models, and improving the balance between precision, comfort, and usability to enhance haptic experiences across virtual reality, assistive technology, and so on.

\subsubsection{Application}
Electrotactile feedback has demonstrated significant versatility across various applications, ranging from skill acquisition and accessibility to teleoperation, notifications, and immersive VR/AR experiences. Electrical stimulation haptic feedback has been effective in improving learning efficiency, enhancing navigation cues, and modulating cognitive load. In assistive technologies, electrotactile interfaces have been used to aid braille recognition, facilitate sign language learning, and improve balance control, underscoring their potential for accessibility and training \cite{Lin_2022, Faltaous_2022, Shindo_2021}.

Beyond training, electrotactile feedback enhances realism in virtual and teleoperated interactions by simulating contact forces, textures, and dynamic affordances \cite{Trinitatova_2022, Ishimaru_2020}. It has also been leveraged for asymmetric interactions and body ownership manipulation, allowing users to partially "surrender" control to external forces, thereby increasing immersion in interactive experiences \cite{Lopes_2015Prop, Patibanda_2023Auto, Patibanda_2023Fused}. Additionally, electrotactile stimulation has proven effective for notifications and navigation, particularly in private alerts and spatial guidance, offering an alternative to visual and auditory cues \cite{Stanke_2023, Pfeiffer_2015Cruise}. In VR and AR environments, electrotactile cues further improve haptic realism, enabling simulated object interactions, dynamic locomotion, and novel hands-free interfaces, expanding the boundaries of immersive experiences \cite{Mukashev_2023, Um_2024}. As technology advances, future research should focus on enhancing comfort, refining stimulation parameters, and integrating electrotactile feedback seamlessly into everyday interactive systems to maximize its usability and impact.

\subsection{Challenges and Limitations}
\label{sec:limitations}

Despite its potential, electrotactile feedback faces several challenges that limit its effectiveness and scalability. These challenges range from device standardization and subjective perception reliance to issues of skin impedance variability, user discomfort, and complex calibration requirements. Addressing these limitations is crucial for advancing electrotactile technology toward broader adoption and improved usability.

\subsubsection{Lack of standardization in electrotactile devices} Electrotactile feedback systems present significant hardware variability, posing challenges for cross-study comparisons and impeding efforts toward standardization. Variability in electrode materials and electrode arrangements often results in inconsistencies across experimental outcomes. Consequently, researchers are compelled to address these hardware discrepancies before conducting meaningful investigations into tactile perception. The absence of a standardized framework further complicates the establishment of consistent benchmarks for evaluating the effectiveness of electrotactile feedback. Kourtesis et al. \cite{kourtesis2022electrotactile} argue that the current limitations of electrotactile devices, particularly their lack of compatibility with common electronic devices such as smartphones, hinder their potential for portable applications. However, we posit that universality should not hinge on compatibility with existing devices but rather on the development of compact, standardized systems. Emerging trends suggest that multimodal electrotactile feedback represents a promising direction for the field \cite{Yem_2017, DrAlonzo_2014, Suga_2024, Suga_2023, Tanaka_2024, Ushiyama_2023}.

\subsubsection{Reliance on subjective perception} Current research primarily relies on psychophysical experiments to interpret tactile sensations, with limited objective evaluation methods available. While subjective reports provide valuable insights into user experience, they introduce variability and post-hoc interpretation biases, making electrotactile perception difficult to predict or replicate. The absence of real-time, quantitative assessment models restricts the ability to develop adaptive and automated haptic feedback systems, ultimately slowing progress in tactile rendering techniques.

\subsubsection{Limited and open-loop tactile rendering} Electrotactile feedback remains predominantly open-loop, with predefined stimulation parameters that do not dynamically adapt to user responses. Most studies focus on device-specific characteristics and parameter tuning rather than real-time adaptation. As a result, tactile rendering is based on prior experimental data rather than personalized feedback loops. This limitation prevents electrotactile systems from adjusting to individual users' skin properties, sensory thresholds, or environmental changes, reducing the reliability of the feedback experience.

\subsubsection{Variability in skin impedance} Skin impedance, which varies due to moisture levels, thickness, and electrode positioning, significantly impacts the perceived intensity and consistency of electrotactile feedback. These physiological variations create inconsistent user experiences, with some users perceiving stronger or weaker sensations than others under identical stimulation settings. Due to the lack of real-time compensation mechanisms, variations in skin impedance lead to reduced precision and repeatability, making it difficult to maintain stable haptic feedback across different users and conditions.

\subsubsection{Complex calibration requirements} Effective electrotactile feedback often requires individualized calibration, adding complexity to system implementation. Differences in skin impedance, sensitivity, and perception thresholds mean that one-size-fits-all settings are ineffective. Current calibration processes involve manual adjustments, making large-scale deployment impractical. Without automated and adaptive calibration techniques, electrotactile systems remain difficult to integrate into consumer devices, medical applications, and everyday wearable technologies.

\subsubsection{The demographic representation of participants is limited} Among the reviewed studies, only one specifically examined the perception of electrical stimulation in older adults, while the rest did not explore age-related differences in electrotactile perception. The majority of user populations studied fall within the 20–35 age range, highlighting a lack of investigation into broader demographic groups, including children, middle-aged adults, and the elderly. This gap in research limits the generalizability of findings and underscores the need for more inclusive studies to better understand electrotactile perception across diverse age groups.

\subsection{Future Research Directions}
Electrotactile feedback holds significant potential for advancing haptic interactions, yet several key areas require further exploration to enhance realism, adaptability, and scalability. Future research should focus on multimodal integration, high-resolution feedback devices, adaptive stimulation algorithms, and full-body electrotactile feedback systems to bridge existing gaps and improve user experiences across various applications.

\subsubsection{Multimodal feedback as the future standard}
A substantial body of research demonstrates that multimodal feedback surpasses unimodal tactile stimulation in terms of both realism and effectiveness \cite{gehrke2019detecting, Kourtesis_2022, Huang_2023, DrAlonzo_2014, Suga_2023, Suga_2024, Lee_2024}. Future electrotactile feedback systems should prioritize hybrid approaches that combine electrotactile stimulation with vibrotactile or force feedback to enhance user immersion and overall system performance.

Meanwhile, due to the lack of a unified solution for electrotactile feedback, the development of complementary software, such as tactile rendering, remains in its early stages. Current research primarily focuses on exploring users' perceptual abilities (often in a single dimension), with limited attention given to the complex tactile rendering of virtual objects. If multimodal devices can be standardized, they will provide a more stable hardware foundation for future researchers, enabling the full potential of complex tactile rendering to be explored and realized.

\subsubsection{High-resolution and scalable feedback devices} As user expectations for high-resolution tactile perception continue to grow, electrotactile devices must adapt to the complexity of the human body and support more sophisticated rendering capabilities. Recent studies suggest that electrical stimulation matrices are better suited for rendering complex information, including spatial and temporal patterns \cite{Ushiyama_2023, blondin2021perception, Dosen_2017, Isakovic_2022, Tanaka_2023, Rahimi_2019Adp, jingu2023double, Zhang_2022, Lin_2022, Franceschi_2017, Gholinezhad_2022, Rahimi_2019Dyn}. Future electrotactile feedback systems should prioritize the development of high-resolution devices to meet the increasing demands for more complex tactile feedback. However, as the complexity of electrode matrices increases, the reliance on microcontroller resources also grows. Therefore, while ensuring high resolution, efforts should also focus on circuit design optimization to reduce system complexity, enabling electrotactile feedback systems to become smaller, lighter, and more portable.

\subsubsection{Adaptive stimulation algorithms for personalized feedback} Current electrotactile feedback systems lack objective biological markers and AI-driven haptic rendering techniques, limiting their ability to provide personalized and real-time adaptive feedback. Future research should explore closed-loop haptic rendering methods that utilize physiological data (e.g., electromyography, impedance monitoring, and real-time sensory feedback) to adjust electrotactile parameters based on individual user responses dynamically \cite{gehrke2019detecting, Rahimi_2019Non}. This will contribute to enabling automated calibration and closed-loop control of electrotactile feedback systems. With the implementation of automated calibration and closed-loop control, electrotactile feedback systems will become more intelligent. At the same time, closed-loop control combined with AI will make tactile rendering possible (as no existing research has yet utilized AI algorithms for tactile rendering). This advancement will pave the way for truly experiencing the metaverse through tactile perception.

\subsubsection{Full-Body electrotactile feedback systems} Despite significant advancements in localized electrotactile interfaces, research on comprehensive full-body haptic systems remains limited (see \autoref{tab:papers}). Most current studies focus on applications involving the hands, wrists, and fingertips, which are critical for fine tactile interactions. However, tactile perception across other regions of the body has been largely underexplored. This limitation is particularly significant as the vision of the metaverse and immersive virtual environments calls for full-body haptic engagement. To meet these demands, it is essential to investigate regional differences in electrotactile sensitivity and develop whole-body wearable electrotactile systems capable of delivering consistent and realistic tactile feedback. 

The distinct distribution of receptors and variations in skin impedance across different body regions result in unique perceptual capabilities, which add a layer of complexity to the design of electrotactile systems. This complexity introduces specific requirements for such devices, particularly in terms of stretchability, comfort, and ease of use, especially for wearable applications that need to conform to the body's shape. While current research emphasizes compact and lightweight designs for localized applications, there has been relatively little exploration of large-area, full-body tactile devices. Addressing these challenges presents significant opportunities for future research, including the development of scalable, flexible, and wearable electrotactile systems designed to accommodate the diverse tactile needs of the human body. 

\section{Conclusion}
This survey provides a comprehensive review of 110 studies on electrotactile feedback in human-computer interaction, examining cutting-edge haptic feedback devices, user perception of electrotactile stimulation, comparisons with other feedback modalities, and its diverse applications. Our analysis underscores the growing role of electrotactile feedback in enhancing haptic experiences, yet challenges remain in device standardization, perception variability, and real-time adaptive rendering.

% \section*{Acknowledgments}
% This research was supported in part by a grant from the Guangzhou Municipal Nansha District Science and Technology Bureau under Contract No.2022ZD01 and the MetaHKUST project from the Hong Kong University of Science and Technology (Guangzhou).

%%
%% The next two lines define the bibliography style to be used, and
%% the bibliography file.
\bibliographystyle{IEEEtran}
\bibliography{reference}

%%
%% If your work has an appendix, this is the place to put it.
% \appendix
% \section{appendix}

\end{document}